\documentclass[]{aa}

\usepackage{aas_macros}
\usepackage{graphicx,txfonts,listings}

\usepackage[colorlinks]{hyperref}
\hypersetup{
    citecolor = {blue},
}

\usepackage{xcolor}

\definecolor{codegreen}{rgb}{0,0.6,0}
\definecolor{codegray}{rgb}{0.5,0.5,0.5}
\definecolor{codepurple}{rgb}{0.58,0,0.82}
\definecolor{backcolour}{rgb}{0.95,0.95,0.92}

\lstdefinestyle{customstyle}{
    backgroundcolor=\color{backcolour},   
    commentstyle=\color{codegreen},
    keywordstyle=\color{magenta},
    numberstyle=\tiny\color{codegray},
    stringstyle=\color{codepurple},
    basicstyle=\ttfamily\footnotesize,
    breakatwhitespace=false,
    xleftmargin=10pt,
    xrightmargin=10pt,
    breaklines=true,                 
    captionpos=b,                    
    keepspaces=true,                 
    numbers=left,                    
    numbersep=5pt,                  
    showspaces=false,                
    showstringspaces=false,
    showtabs=false,                  
    tabsize=2
}

\usepackage{tikz}
\usetikzlibrary{shapes,arrows,positioning,calc,angles}

\newcommand{\dd}{{\rm d}}

\newcommand{\bvec}[1]{\mathbf{#1}}

\newcommand{\HERMES}{{\tt HERMES}}
\newcommand{\GammaSky}{{\tt GammaSky}}
\newcommand{\GALPROP}{{\tt GALPROP}}
\newcommand{\CRPROPA}{{\tt CRPropa3}}
\newcommand{\HEALPIX}{{\tt HEALPix}}
\newcommand{\DRAGON}{{\tt DRAGON}}
\newcommand{\HAMMURABI}{{\tt HAMMURABI}}
\newcommand{\CLUMPY}{{\tt CLUMPY}}
\newcommand{\PICARD}{{\tt PICARD}}

\begin{document}

\author{A. Dundovic\inst{1,2} \and C. Evoli\inst{1,2} \and D. Gaggero\inst{3} \and D. Grasso\inst{4}}

\institute{Gran Sasso Science Institute, Viale Francesco Crispi 7, I-67100 L'Aquila, Italy
\and
INFN Laboratori Nazionali del Gran Sasso (LNGS), I-67100 Assergi, L'Aquila, Italy
\and
Instituto de F\'isica Te\'orica UAM-CSIC, Campus de Cantoblanco, E-28049 Madrid, Spain
\and
INFN Sezione di Pisa, Polo Fibonacci, Largo B. Pontecorvo 3, I-56127 Pisa, Italy
}

\title{Simulating the Galactic multi-messenger emissions with HERMES}

\abstract{The study of nonthermal processes such as synchrotron emission, inverse Compton scattering, bremsstrahlung, and pion production is crucial to understanding the properties of the Galactic cosmic-ray population, to shed light on their origin and confinement mechanisms, and to assess the significance of exotic signals possibly associated to new physics.}{We present a public code called \HERMES~which is designed generate sky maps associated to a variety of multi-messenger and multi-wavelength radiative processes, spanning from the radio domain all the way up to high-energy gamma-ray and neutrino production.}{We describe the physical processes under consideration, the code concept and structure, and the user interface, with particular focus on the {\tt python}-based interactive mode. In particular, present the modular and flexible design that allows the user to easily extend the numerical package according to their needs.}{In order to demonstrate the capabilities of the code, we describe the details of a comprehensive set of sky maps and spectra associated to all physical processes included in the code. We comment in particular on the radio, gamma-ray, and neutrino maps, and mention the possibility of studying signals stemming from dark matter annihilation.}{\HERMES~can be successfully applied to constrain the properties of the Galactic cosmic-ray population, improve our understanding of the diffuse Galactic radio, gamma-ray, and neutrino emission, and search for signals associated to particle dark matter annihilation or decay.}


\maketitle

\section{Introduction}
\label{sec:intro}

The Milky Way galaxy has been recognized over recent decades as a bright broad-band source of diffuse nonthermal multi-messenger radiation, from MHz radio waves all the way up to multi-TeV gamma-rays and neutrinos. 
A coherent modeling of the different components that contribute to these emissions, and confrontation with available data, is required to reach a comprehensive understanding of the nature of this radiation.

In the low-frequency domain, the radio and microwave emissions trace several key interactions between the Galactic cosmic-ray (CR) leptons and the interstellar medium (ISM): in particular, the synchrotron component originates from interactions with the Galactic magnetic field.
For magnetic field intensities of $\mathcal{O}(1)\,\mu$G, comparable with the typical values inferred in the Milky Way, and for CR leptons of [GeV -- TeV] energies, the synchrotron radiation falls in the [MHz -- GHz] frequency range. 
Therefore, the study of the radio maps via extensive surveys, together with a variety of complementary observables  --- including, for instance, the deflection of ultra-high energy cosmic rays (UHECRs), and the study of the Faraday rotation of polarized sources --- allows the distribution of high-energy CR leptons and the structure of the Galactic magnetic field to be constrained and modeled. 
This line of research has been developed over the last and current century and has led to the identification  of large coherent magnetic structures in the Galaxy with ever improving accuracy~\citep{Beuermann1985aa,Men2008aa,Sun2008aa,Sun2010raa,Jaffe2010mnras,Jaffe2011mnras,Fauvet2011aa,Jansson2012apj,Unger2017icrc,Hutschenreuter2021arxiv}. 
As far as CR studies are concerned, the analysis of the Galactic synchrotron maps revealed useful information about the spectral properties of the diffuse population of relativistic leptons and their distribution in the Galactic halo~\citep{Strong2011aa,DiBernardo2013jcap}.
Within the same frequency range,  free–free emission is produced by thermal electrons scattering off ions in the ionized component of the ISM. 
This process is particularly relevant above a few GHz in the Galactic plane in emission and below $\simeq 300$ MHz in absorption. 

On the other end of the electromagnetic spectrum, gamma-ray and neutrino astronomy are younger disciplines. 
The first detection of a diffuse signal above 50 MeV from the Galactic plane dates back to the pioneering observations performed by the OSO-3 Satellite in 1967-1968 \citep{Kraushaar1972apj} followed by SAS-2~\citep{Kniffen1973apj}, COS-B~\citep{Lebrun1982aa}, and EGRET~\citep{Hunter1997apj}. 

Over recent years, the AGILE telescope~\citep{Tavani2009aa} and the FERMI Gamma-Ray Space Observatory~\citep{Atwood2009apj} have enriched the catalog of Galactic and extra-galactic gamma-ray sources,  also providing detailed full-sky maps of the diffuse emission up to 1 TeV with an angular resolution reaching $\simeq 0.1^\circ$.
At even larger energies, a crucial role is played by ground-based Atmospheric Cerenkov Telescopes (ACTs) such as HESS~\citep{Aharonian2000head}, VERITAS~\citep{Weekes2002aph}, MAGIC~\citep{MAGIC2005}, and Water Cerenkov Telescopes (WCTs) such as MILAGRO~\citep{Milagro2001} and HAWC~\citep{Westerhoff2014adspr}.
The ACTs feature a limited field of view, but they can probe the diffuse emission with higher sensitivity and resolution in limited but very compelling regions like the Galactic center (GC) or several  star-forming regions.

Thanks to the accuracy of the current measurements provided by satellite experiments, new unexpected features have emerged in the spatial properties of the diffuse CR sea~\citep{Gaggero2015prd,Acero2016apjs,Yang2016prd,Pothast2018jcap}, with profound implications for the underlying micro-physics of CR transport~\citep{Cerri2017jcap,Recchia2016mnras}, and phenomenological implications for the interpretation of very-high-energy gamma-ray and neutrino data~\citep{Gaggero2015apj,Neronov2016aph,Pagliaroli2016jcap,Gaggero2017prl,Cataldo2019jcap}. 
Moreover, ground-based experiments have revealed the first signs of high-energy diffuse emission from bright regions near the GC in the TeV domain. 

At these energies, gamma-rays are partially attenuated because of the presence of the interstellar radiation fields, and therefore high-energy neutrinos offer a valuable complementary probe of the hadronic components of CRs. 
High-energy ($E \gtrsim 10$ TeV) neutrino astronomy has recently come online with the detection of several neutrinos of unambiguous astrophysical origin by the IceCube observatory at the South Pole (see~\citealt{Ahlers2018prnp} for a recent review). 
Recently, the experiment ANTARES in the Mediterranean sea  also joined the effort for the detection of astrophysical neutrinos~\citep{Albert2018apj}.  
In particular, a combined analysis of the Galactic plane performed by the two experiments provided the first hint of a component of the emission caused by the galactic CR population~\citep{Albert2017prd,Albert2018apjb}.

Besides being a relevant problem {\it per se}, the study of the diffuse nonthermal emission from the Galaxy has an important link to the quest for the elusive dark matter (DM) component that permeates the Universe. 
Under the hypothesis that this substance is made of weakly interacting massive particles (WIMPs), a gamma-ray signal may be expected from the annihilation or the decay of those particles in the DM halo that embeds the Milky Way galaxy. 
The search for such a signal in DM-rich regions, in particular the inner Galaxy, is a challenging task that requires accurate modeling of the astrophysical processes and advanced data analysis, and is clearly hampered by the complexity of the astrophysical processes outlined above~(see~\citealt{Gaggero2018review} for a review). 

The future is extremely promising for both extremities of the electromagnetic spectrum and for the neutrino channel, opening unprecedented scientific opportunities for the coming years.
In the radio domain, the Square Kilometer Array (SKA) will operate over a wide range of frequencies, and its size will typically improve the sensitivity by a factor of $\sim$ 50 compared to previous instruments~\citep{SKA2020}. 
In the very-high-energy gamma-ray domain, the Cerenkov Telescope Array (CTA)~\citep{CTA2019} and the Large High Altitude Air Shower Observatory (LHAASO)~\citep{LHAASO2019} will provide a giant leap in sensitivity in both the Norther and Southern hemisphere. 

In particular, CTA will characterize high-energy emitters and the diffuse emission along the Galactic plane, helping to improve our understanding of CR acceleration up to PeV energies and of the properties of the CR diffuse sea above the TeV.  It will also help in advancing the search for a DM signal from multi-TeV WIMPs with an annihilation cross-section as low as the one usually associated to thermal production in the early Universe. 
Concerning neutrinos, IceCube (and future extensions) will soon be assisted by~KM3NeT~\citep{KM3NET2016} which, being located in the Mediterranean sea, will allow a better and more accurate view of the inner Galactic plane. 

The plethora of existing data discussed above and the impressive increase in sensitivity expected from future experimental facilities motivates a complementary effort on the theoretical side.
Several public codes have been released over the years to model at least some of those data sets.
Concerning the radio window, the \HAMMURABI\footnote{\url{https://github.com/hammurabi-dev/hammurabiX}}~code was specifically designed to simulate the Galactic synchrotron emission and Faraday rotation~\citep{Waelkens2009aa}. 
In particular, its latest release entails an accurate description of the turbulent field and the calculation of the polarized emission~\citep{Wang2020apjs}.
In the gamma-ray range, \GALPROP\footnote{\url{https://galprop.stanford.edu}}~has been widely used to reproduce the general features of the interstellar gamma-ray emission over the whole sky~\citep{Strong2009galprop,Moskalenko2019icrc}.
\GALPROP~is designed to simulate Galactic CR propagation and associated diffuse emissions  simultaneously. 
The code computes the gas-related gamma-ray intensities using the CR flux computed beforehand and the column densities of HI and H2 for Galactocentric annuli based on 21-cm and CO surveys as described in~\cite{Strong2004apj}. 
The inverse Compton scattering is treated using the formalism for an anisotropic radiation field developed by~\cite{Moskalenko2000apj}.
A new version of the code\footnote{\url{https://gitlab.mpcdf.mpg.de/aws/galprop}} includes detailed calculation of the synchrotron (also in polarization) and free--free emission, and provides the possibility to compare different recent models of the Galactic magnetic field.
A semi-analytical approach to modeling the main components of the diffuse gamma-ray emission of the Galaxy has also been proposed ~\citep{Delahaye2011aa}. 
The \CLUMPY\footnote{\url{https://gitlab.com/clumpy/CLUMPY}}~code provides a comprehensive framework with which to compute indirect gamma-ray and $\nu$ signals from the Galactic DM annihilation or decay extending to the extragalactic scales and to model the contribution from substructures~\citep{Hutten2019cophc}.
The \PICARD\footnote{\url{https://astro-staff.uibk.ac.at/~kissmrbu/Picard.html}}~code ---which introduced a new and fast numerical treatment of cosmic-ray transport in 3D--- was recently applied to model the diffuse gamma-ray emission of the Galaxy between 100 MeV and 100 TeV ~\citep{Kissmann2014aph,Kissmann2015aph}.
Finally, in the context of gamma-ray analyses, we mention the D$^3$PO framework \citep{Selig2015aa}. This method is designed to remove the shot noise, deconvolve the instrumental response, and to provide estimates for the different flux components separately by means of a Bayesian inference technique.

Here, we present the simulation framework \HERMES~(High-Energy Radiative MESsengers) which joins these efforts and can be used to constrain the properties of the Galactic CR population, increasing our understanding of the radio and gamma-ray diffuse Galactic emission.

\HERMES~is designed for efficient development of astrophysical predictions for Galactic diffuse emissions. 
Users can assemble modules of the relevant quantities describing the Galactic environments (magnetic fields, radiation fields, CR densities, etc.), including their own modules, and receive as an output full-sky radio, gamma-ray, or neutrino maps associated to the following nonthermal radiative processes:  (a) radio emission due to free--free scattering of thermal electrons onto the ionized component of the ISM; (b) synchrotron emission in the MHz - GHz domain (eventually including the absorption by free--free); (c) gamma-ray emission by Inverse Compton scattering of the diffuse low-energy background photons; (d) gamma-ray and neutrino emission by pion-decay (eventually including the gamma-ray absorption by pair production); (e) gamma-ray emission by bremsstrahlung; (f) and gamma-ray and neutrino emission by DM annihilations in the dark matter halo. 
We also implement the calculation of the Galactic faraday rotations and dispersion measures as they can be used to constrain the distribution of the magnetic fields and free electron density independently of the CR density.

From a technical point of view, \HERMES~features a modular {\tt C++} structure combined with a {\tt Python} interface for user-friendliness (specifically in the I/O). This combination takes advantage of the popularity of {\tt Python}~\citep{Momcheva2015arxiv} and at the same time benefits from the fast computational performance of a {\tt C++} code~\citep{Portegies2020natas}.

The paper is structured as follows. 
In section 2 we briefly summarize the physics of the processes involved. 
In section 3, the code structure of the program is presented. 
The capabilities for simulating diffuse emissions are demonstrated in section 4 in a few examples where we also introduce a selected choice of up-to-date models of the CR distributions, Galactic magnetic fields, Galactic gas distribution (ionized, atomic, and molecular), and interstellar radiation fields (ISRFs) that are included in the program.
Finally, results are summarized in section 5.

The program source code is licensed under the GNU General Public License v3 and is publicly available\footnote{\url{https://github.com/cosmicrays/hermes}} together with installation instructions and examples\footnote{\url{https://github.com/cosmicrays/hermes-examples}}.
Questions and comments can be submitted to the ticketing system\footnote{\url{ https://github.com/cosmicrays/hermes/issues}}.

\section{Physical processes included in the code}
\label{sec:physics}

In this section, we describe the physical processes that contribute predominantly to the gamma-ray ($E_\gamma \gtrsim 100$~MeV) and radio ($\nu$ in MHz $\div$ 100 GHz) diffuse emissions from our Galaxy and their implementation in the \HERMES~code.

The local emissivities (i.e., production rate per unit volume) are expressed as $\epsilon_E$ or $\epsilon_\nu$ depending on whether the emissivity is expressed as differential in energy or in frequency, respectively. 

The differential intensity (differential flux per unit angle) as a function of direction in the sky is obtained by integrating the emissivity over the line of sight (LOS) distance $s$:
\begin{equation}
I(x, l, b)= \frac{1}{4\pi} \int_0^\infty \! {\rm d}s \, \epsilon_x(\bvec r, x)
,\end{equation}
where the longitude $l$ and the latitude $b$ constitute a spherical coordinate system centered on the Sun and $x$ is either the gamma-ray energy $E_\gamma$ or the radio frequency $\nu$.

The conversion between $(l,b,s)$ and the cartesian coordinate system centered at the Galactic center $\bvec{r} = (x,y,z)$ is given by:
\begin{equation}
z = s \sin b \, , \quad x = s \cos b \cos l - r_\odot \, , \quad y = s \cos b \sin l.
\end{equation}

In the code we adopt SI units. 

\subsection{Faraday rotation}

The polarization angle of an electromagnetic wave is rotated when crossing a magnetized plasma, an effect known as Faraday rotation~\cite[see e.g.,][]{Rybicki1979book}. 
Faraday effects on pulsar and extragalactic radio source data have been used to model the magnetic fields of the Milky Way~\citep{Pshirkov2011apj}. 
In particular, the Faraday rotation towards pulsars are due to the ISM in the direction of the inner Galaxy, while data on extragalactic sources provide information about the magnetic fields over much of the regions above and below the Galactic plane.

The rotation measure (RM) quantifies the rate of change of the polarization angle $\chi$ as expressed by the formula:
\begin{equation}
\chi = \text{RM} \, \lambda^2 + \chi_0 
,\end{equation}
where $\chi_0$ is the original (or intrinsic) angle at the polarized source and $\lambda$ is the observation wavelength.

The RM is given by the integral of the magnetic field component {along} the 
LOS $B_{\|}$ weighted by the thermal electron density $n_e$~\citep{Jackson1975book}:
\begin{equation}
{\rm RM}(l,b) = \frac{q^3}{8 \pi^2 \epsilon_0 m_e^2 c^3} \int_0^\infty \! \dd s \, n_{\rm e}(\bvec{r}) B_\| (\bvec{r}) 
\label{eq:rotationmeasure}
,\end{equation}
where $q$ is the elementary charge, $c$ the speed of light in vacuum, $m_e$ is the electron mass, and $\epsilon_0$ is the electric constant.
The Faraday rotation maps are conveniently expressed in [rad~m$^{-2}$].

Another useful quantity in radioastronomy is the dispersion measure 
which corresponds to the integrated column density of free electrons and is calculated using:
\begin{equation}
{\rm DM}(l,b) = \int_0^\infty \dd s \, n_e(\bvec r)
\label{eq:dispersionmeasure}
.\end{equation}
Dispersion measure is often given in units of [pc~cm$^{-3}$].

\subsection{Free--free emission and absorption}
\label{sec:freefree}

Free--free emission is produced by free (thermal) electrons scattering off ions in the ionized component of the ISM.
This radiation contributes to the unpolarized radio diffuse emission at frequencies above a few GHz in the Galactic plane where it is produced in the gas layer ionized by radiation from recently formed stars.
Free--free transitions of ionized hydrogen are also relevant in absorption of synchrotron emission.
This is important at frequencies below $\sim$GHz and predominantly along the Galactic plane, but also slightly affects the synchrotron radiation observed at higher frequencies.

For hydrogen gas, the free--free emissivity by neutral plasma (namely when the ion density is equal to the free-electron density: $n_{\rm HII} = n_{\rm e}$) at a given observed frequency $\nu$ is taken as in~\cite{Longair2011book}:
\begin{equation}
\epsilon_\nu  (\nu, T_e, \bvec r) = 
\frac{1}{3\pi^2} \sqrt{\frac{\pi}{6}} 
\frac{q^6}{\epsilon_0^3 c^3 m_e^2} \sqrt{\frac{m_e}{k_B T_e}} g(\nu, T_e) \, n_e(\bvec r)^2 \exp\left(-\frac{h\nu}{k_B T_e}\right) 
,\end{equation}
where $k_B$ is Boltzmann constant and $T_e \sim 10^4$~K is the average temperature of the free electrons in the ISM.

The Gaunt factor at radio wavelengths can be approximated by:
\begin{equation}
g(\nu, T_e) = \frac{\sqrt{3}}{2\pi} \left[ \ln\left( \frac{128 \epsilon_0^2 k_B^3 T_e^3}{m_e q^4 \nu^2} \right) - \Gamma^{1/2}\right]
,\end{equation}
where $\Gamma \simeq 0.577$.

From the emissivity, we additionally compute the absorption coefficient $\alpha_\nu^{\rm ff}$ (i.e., the probability of absorption per unit distance) as in~\cite{Rybicki1979book}:
\begin{equation}
\alpha_\nu^{\rm ff} (\nu, T_e, \bvec r)  = \frac{\epsilon_\nu(\nu, T_e, \bvec r) }{4 \pi B_\nu(\nu, T_e)} 
,\end{equation}
where $B_\nu$ is the Planck spectrum of the black-body radiation at the temperature $T_e$. 

\subsection{Synchrotron emission}

The radio continuum emission of the Milky Way below $\sim$100 GHz mostly originates from the synchrotron process and hence observations of synchrotron intensity and spectral index provide stringent constraints on the interstellar electron spectrum and on the galactic magnetic field models~\citep{Jansson2012apj,Strong2011aa,DiBernardo2013jcap,Vittino2019prd}. 
At low frequencies, absorption of synchrotron emission by thermal gas (free--free absorption) can become sizeable and must be taken into account.

The emissivity of a population of relativistic electrons is given by~\cite{Longair2011book}:
\begin{equation}
\epsilon_\nu(\bvec r, \nu) = 
\frac{4\pi}{c} \int \! \dd E \, \Phi_e(E, \bvec r) j_\nu(E, \bvec r, \nu)
,\end{equation}{}
where $\Phi_e$ is the differential spectrum of relativistic electrons of energy $E$, and $j_\nu$ is the total emissivity of a single electron at a given observational frequency $\nu,$ and depends on the magnetic field projected in the direction perpendicular to the LOS $B_\perp$ as:
\begin{equation}\label{eq:singleemissivity}
j_\nu = \frac{\sqrt{3}q^3 B_\perp}{8\pi^2 \epsilon_0 c m_e} F\left(\frac{\nu}{\nu_c}\right)
.\end{equation}

In Eq.~\ref{eq:singleemissivity} we define the critical frequency $\nu_c$ as:
\begin{equation}
\nu_c(E, \bvec r) = \frac{3}{4\pi} \gamma_e^2(E) \frac{q B_\perp(\bvec r)}{m_e}
,\end{equation}
where $\gamma_e$ is the electron Lorentz factor, and $F(x)$ is defined in terms of the Bessel function $K_{5/3}$:
\begin{equation}
F(x) = x \int_x^\infty \! \dd x' \, K_{5/3}(x')
.\end{equation}

In order to account for free--free absorption, the LOS integration to compute the synchrotron intensity as a function of the LOS direction becomes:
\begin{equation}
I_\nu(\nu,l,b) = \frac{1}{4\pi} \int_0^\infty \dd s \,  A_\nu(\nu, \bvec r) \epsilon_\nu(\bvec r, \nu)  
,\end{equation}
where $A_\nu = \exp\left(- \int_0^s \! ds' \, \alpha_\nu^{\rm ff}(\bvec r') \right)$ is the absorption function which depends on the absorption coefficient $\alpha_\nu^{\rm ff}$ that we introduce in~\S~\ref{sec:freefree}.

As radio emission is often associated with thermal phenomena, the intensity is traditionally stated in terms of the brightness temperature $T_b$ defined as:
\begin{equation}
T_b(\nu,l,b) = \frac{c^2}{2 \nu^2 k_B} I_\nu(\nu,l,b)
.\end{equation} 

\subsection{Inverse Compton}

The inverse Compton (IC) diffuse emission is produced by CR electrons and positrons scattering on low-energy target photons in the Galaxy as UV, optical, infrared, or microwave background photons (the sum of all the fields form the interstellar radiation field (ISRF)). 
The IC scattering is a major contributor to the diffuse Galactic emission above $\sim$~MeV, and becomes the dominant process at high galactic latitudes because the intensity of gamma radiation from interactions between CRs and gas is greatly reduced~\citep{Acero2016apjs}.

The differential cross-section for producing gamma-rays of energy $E_\gamma$ by a high-energy electron $E_e$ scattering on a low-energy photon $E_{\rm ph}$ is~\citet{Blumenthal1970rvmp}:
\begin{multline}
\frac{\dd\sigma_{\rm IC}}{\dd E_\gamma}( E_e, E_{\rm ph}, E_\gamma) = \\
\frac{3 \sigma_{\rm T}}{4 E_{\rm ph} \gamma_e^2} \left[ 2q \ln{q} + (1 + 2q)(1-q) + \frac{(p q)^2 (1-q)}{2(1+p q)}  \right], 
\end{multline}
where $\sigma_{\rm T}$~is the Thomson cross-section,
\begin{equation}
p = \frac{4 E_{\rm ph} E_e}{(m_e c^2)^2}
,\end{equation}
and
\begin{equation}
q = \frac{E_\gamma}{4 E_{\rm ph} \gamma_e^2 \displaystyle{\left(1-\frac{E_\gamma}{E_e}\right)}} \,\,\, \text{in the range} \,\,\, \frac{1}{4\gamma_e^2} < q \le 1 
.\end{equation}

The gamma-ray emissivity is obtained by integrating over the ISRF and over the CR spectrum as follows:
\begin{multline}
\label{eq:ICemisssivity}
\epsilon_{\rm E}(E_\gamma, \bvec r) = 
4 \pi \int \! \dd E_{\rm ph} \, \frac{\dd n_{\rm ph}}{\dd E_{\rm ph}}(E_{\rm ph}, \bvec r) \, \\
\int \! \dd E_{\rm e} \, \frac{\dd \sigma_{\rm IC}}{\dd E_\gamma} (E_{\rm e}, E_{\rm ph}, E_\gamma) \Phi_{\rm e}(E_e, \bvec r), 
\end{multline}
where $n_{\rm ph}$ is the background photon density and $\Phi_{\rm e}$ is the CR (electrons and positrons) differential flux.

In Eq.~\ref{eq:ICemisssivity} we assume isotropic scattering in the sense that both the lepton distribution and the ISRF are assumed isotropic.
The effects of anisotropic scattering on the interstellar radiation field (apart from the cosmic microwave background) have been discussed in~\citet{Moskalenko2000apj,Orlando2021jcap}~\footnote{See also \url{https://gitlab.mpcdf.mpg.de/aws/stellarics}}.

\subsection{Gamma-ray emission by pion decay}\label{sec:piondecay}

Gamma rays are produced by the decay of neutral pions created in collisions of CR nuclei with interstellar gas atoms.
Neutral pion production is a catastrophic energy loss process for the protons which typically retain about $\sim20$\% of their energy after interaction.
The gamma-ray emission can then be used to trace the CR distribution in the Galaxy and to probe its spectrum far from the Solar System~\citep{Gaggero2015apj,Acero2016apjs,Yang2016prd}.

As the most abundant species in the CR flux are proton and helium, the emissivity from collisions between CR and gas nuclei can be written as:
\begin{multline}
\epsilon_{E}(E_\gamma, \bvec r) = 4 \pi \, n_{\rm H}(\bvec r) \int \dd E \,\left[ \Phi_{\rm H}(E, \bvec r) \left( \frac{\dd\sigma_{\rm p-p}}{\dd E_\gamma} +  \ f_{\rm He} \frac{\dd\sigma_{\rm He-p}}{\dd E_\gamma} \right) + \right. \\ \left. \Phi_{\rm He}(E, \bvec r) \left( \frac{\dd \sigma_{\rm p-He}}{\dd E_\gamma} + f_{\rm He} \frac{\dd\sigma_{\rm He-He}}{\dd E_\gamma} \right) \right], \label{eq:pi0emissivity}
\end{multline}
where $n_{\rm H}$ is the interstellar hydrogen density, $\Phi_i$ is the CR differential flux of the CR species $i$ as a function of the kinetic energy per nucleon $E,$ and $\dd \sigma_{\rm i-j}/\dd E_\gamma$ is the differential cross-section of secondary gamma-ray production in $i-j$ interactions. 
We assume here that the ISM gas is a mixture of hydrogen and helium nuclei with uniform density ratio $f_{\rm He} = 0.1$.

The interstellar hydrogen is predominantly made of neutral (HI) and molecular hydrogen (H$_2$). 
These gas components, especially H$_2$ which is mostly concentrated in dense clouds, follow complex geometrical structures which cannot be directly inferred from observations. 
In fact, a detailed knowledge of the gas kinematics is required in order to translate spectroscopic data into three-dimensional distributions. 
Some models for the HI and H$_2$ 3D spatial distributions have been developed either under the form of analytical expressions or as tables where the gas density can be read at given positions~\citep{Nakanishi2003pasj,Ferriere2007aa,Pohl2008apj,Mertsch2020arxiv}. 
These models can be easily used as input in \HERMES~to compute the gamma-ray emissivity as a function of position and finally the intensity along a given LOS. 

An alternative strategy originally devised by the authors of the \GALPROP~code \citep{Strong2000apj,Strong2009galprop} makes direct use of the column density skymaps derived from radio measurements.
Moreover, column
density maps have been obtained in the form of  ``rings''  corresponding
to different intervals of the galactocentric radius by combing sky surveys of the 21 cm emission of HI and surveys of CO spectral lines -- tracing H$_2$ -- with Galactic rotation curves. 

We can then obtain the total gas column density $N_{\rm H}$ within the $i$-th ring and as a function of $(l,b)$ as:
\begin{equation}
N_{\rm H}^i(l,b) = N^i_{\rm HI}(l,b) + 2 N^i_{{\rm H}_2}(l,b) = N^i_{\rm HI}(l,b) + 2 X^i_{\rm CO} w^i_{\rm CO}(l,b)
,\end{equation}
where $X_{\rm CO}(r)$ is the CO-H$_2$ conversion factor and is given in units of [$10^{20}$~cm$^{-2}$ K$^{-1}$ / (km s$^{-1}$)] and $W_{\rm CO}$ is the velocity-integrated CO line intensity measured in [K km s$^{-1}$]. 

The observed intensity in a given direction can consequently obtained as the sum of the gas in each ring multiplied by the emissivity averaged over the ring:
\begin{equation}\label{eq:ringmodel}
I_\gamma(l,b,E_\gamma) = \frac{1}{4\pi} \sum_i N_{\rm H}^i(l,b) \langle \epsilon_E(\bvec r, E_\gamma) \rangle^i
,\end{equation}
where
\begin{equation}\label{eq:averagemissivity}
\langle \epsilon_E(\bvec r, E_\gamma) \rangle^i = \frac{ \int_0^\infty {\rm d}s \, \epsilon_{E}  (E_\gamma, \bvec{r}) p_{\rm HI}(\bvec{r}) \Theta^i_{\rm in}(\bvec{r})
}{\int_0^\infty {\rm d}s \, p_{\rm HI}(\bvec{r})\Theta^i_{\rm in}(\bvec{r})}
,\end{equation}
and \(p_{\rm HI}\) is a smooth function that describes the gas profile at the Galactic scale, and \(\Theta^j_{\rm in}(\bvec{r})\) is the Theta-function defined as:
\begin{equation}
\Theta^j_{\rm in}(\bvec{r}) = 
    \begin{cases}
    1\quad {\rm if}\quad \bvec{r} {\rm\ inside\ the\ }i-{\rm th\ ring},\\
    0\quad {\rm if}\quad {\rm elsewhere.}
    \end{cases} 
\end{equation}

\subsection{Bremsstrahlung}

Relativistic electrons and positrons additionally lose their energy in the Galaxy by Bremsstrahlung: radiation is emitted by a lepton passing through the electric field of a particle in the ISM (electron or nucleus).
As for the synchrotron radiation, this mechanism is inversely proportional to the particle mass and is therefore negligible for protons and nuclei.

The differential bremsstrahlung emissivity in the interstellar space is calculated by:
\begin{multline}
\epsilon_{E}(E_\gamma, \bvec r) =  4 \pi \, n_{\rm H}(\bvec r) \int_{E_\gamma}^\infty \dd E \Phi_e(E, \bvec r) \left[ \frac{\dd\sigma_{\rm br}}{\dd E_\gamma}({\rm H}, E_\gamma, E) + \right. \\ \left. f_{\rm He} \frac{\dd\sigma_{\rm br}}{\dd E_\gamma}({\rm He}, E_\gamma, E) \right],
\end{multline}
where $\dd \sigma_{\rm br}/ \dd E_\gamma$ is the differential cross-section for the emission of a photon of energy $E_\gamma$ by an electron or a positron with kinetic energy $E$, and we distinguish between the cases for H and He.

After computing the emissivity, we obtain the intensity as for the pion decay case by summing over the gas rings as in Eq.~\ref{eq:ringmodel}.

\subsection{The attenuation of gamma-rays due to pair production}

Gamma rays in the energy range $E_\gamma \gtrsim 10$~TeV suffer non-negligible absorption during their propagation from the emission point to the Earth. The mechanism that generates the absorption is the creation of electron–positron pairs in photon–photon interactions $\gamma \gamma \rightarrow e^+e^-$, where the target is provided by the ISRF and the CMB photons~\citep{Moskalenko2006apj}.

The cross-section of the process $\gamma \gamma \rightarrow e^+e^-$ is~\citep{Vernetto2016prd}:
\begin{equation}
\sigma_{\gamma\gamma} = \sigma_{\rm T} \frac{3}{16} (1-\beta^2) \left[ 2\beta(\beta^2-2)+(3-\beta^4) \ln \frac{1+\beta}{1-\beta} \right],
\end{equation}
where $\beta = \sqrt{1-\frac{1}{x}}$, $x = \displaystyle{ \frac{E_\gamma E_{\rm ph}}{2 (m_e c^2)^2}(1-\cos \theta),}$ and $\theta$ is the angle between the directions of the interacting photons.

The absorption probability per unit path length (or absorption coefficient) for a gamma-ray of energy E$_\gamma$ at the space point $\bvec r$ 
can be calculated integrating the cross-section over the energy and angular distributions of the target photons (which we assume to be isotropic):
\begin{multline}
K_{pp}(E_\gamma, \bvec r) = 2 \int \! \dd\theta \, (1-\cos \theta) \\ \int_{E_{\rm ph}^{\rm min}}^\infty \! \dd E_{\rm ph} \,  n_\gamma(E_{\rm ph}, \bvec r) \sigma_{\gamma \gamma} [\beta(E_\gamma, E_{\rm ph}, \theta)], 
\end{multline}
where $E_{\rm ph}^{\rm min} = 2 m_e^2 c^4/ E_\gamma (1-\cos\theta)$.

The optical depth $\tau$ for photons of energy $E_\gamma$ traveling over the distance $\Delta s$ is obtained by integrating the absorption coefficient $K$ along the LOS:
\begin{equation}
\tau(E_\gamma, \Delta s) = \int_{\Delta s} \! \dd s \, K_{pp}(E_\gamma, \bvec r)
.\end{equation}

The pair-production absorption is believed to be relevant only for gamma-rays with energy much larger than $\sim$TeV and passing near the Galactic Centre. 
%
In these conditions, the Galactic emission is dominated by the $\pi$-decay and therefore the absorption is only implemented for this process.

However, as accounting for the absorption implies numerically evaluating  an additional integral along the LOS for each direction and each ring, which can be computationally expensive, the \HERMES~user can easily turn the calculation of  the absorption  on or
off to speed up the calculation. 
When the absorption is taken into account, the ring-averaged gamma-ray emissivity  is computed as:
\begin{equation}\label{eq:ringmodel_abs}
\langle \epsilon_E(\bvec r, E_\gamma) \rangle^i = \frac{ \int_0^\infty \dd s \, \epsilon_{E}  (E_\gamma, \bvec{r}) p(\bvec{r}) \exp \left[-\tau(E_\gamma, \bvec{r}) \right] \Theta^i_{\rm in}(\bvec{r})
}{\int_0^\infty \dd s \, p(\bvec{r}) \Theta^i_{\rm in}(\bvec{r})} ,\end{equation}
where $i$ is the index of the gas-ring considered.

\subsection{Neutrinos from pion decay}

The diffuse gamma-ray emission of the Galaxy due to the hadronic scattering of the CR sea onto the ISM gas via $\pi^0$ decay is accompanied by a corresponding neutrino emission via the decay of charged pions and muons. 
As gamma rays are also produced by leptonic processes, the possible detection of the Galactic diffuse neutrino emission may then offer a better probe of the CR nuclei population.
Moreover, as opposed to gamma-rays, neutrinos are unattenuated even in the PeV energy range, offering a unique way to probe the primary CR spectrum in the knee region even in very far and opaque regions of the Galaxy. 

To compute the electron and muon neutrino production spectra we follow the same approach as for gamma-rays.
We implement the neutrino emissivity in \HERMES~using the gas-ring model as in Eq.~\ref{eq:ringmodel}.
The cross-sections $\dd \sigma_{\rm CR-p} / \dd E_\nu$ are now computed by summing the contribution of CR-$p$ scattering to $\nu_e$ and $\nu_\mu$ (and their anti-neutrino counterparts).
It can be shown (see e.g., ~\citealt{Palladino2020univ}) that neutrino vacuum oscillations on Galactic distances distribute the total flux among all neutrino flavors almost equally.

\subsection{Dark matter annihilations}
\label{sec:dm}

If the DM consists of weakly interacting massive particles (WIMPs), an important tool for inferring their properties could be the detection of gamma-rays and neutrinos produced by the annihilation products of DM in our Galaxy. 
For WIMPs with masses close to the electroweak scale, $m_\chi \sim 100$~GeV - 1 TeV, the annihilation products are typically found in the  GeV-TeV range.

The gamma-ray emissivity produced in DM annihilations is described by:
\begin{equation}\label{eq:dmemiss}
\epsilon_E(E_\gamma, \bvec r) = \frac{1}{4\pi} \frac{1}{2} \frac{\langle \sigma v \rangle}{m_{\rm \chi}^2}  \frac{\dd N}{\dd E_\gamma} \rho^2_\chi(r)
,\end{equation}
where $m_\chi$ is the mass of the WIMP, and $\rho_\chi$ is the  DM density depending only on the distance to the Galactic Centre $r$. 
The gamma-ray spectrum generated per WIMP annihilation is $\dd N/\dd E_\gamma$, normalized such that its integral over energy is equal to 1. 
The factor `1/2' accounts for the fact that the DM is assumed to be its own antiparticle.
The $\langle \sigma v \rangle$ is the WIMP annihilation cross-section multiplied by the relative velocity of the two WIMPs (averaged over the WIMP velocity distribution), and we assume as benchmark value $\langle \sigma v \rangle = 3 \times 10^{-26}$~cm$^3$/s.
Similarly, the neutrino emissivity is obtained as in Eq.~\ref{eq:dmemiss} with the difference that the neutrino spectrum $\dd N / \dd E_\nu$ has to be computed.

The integral over the LOS determines the angular dependence of the signal and is controlled by the astrophysical distribution of DM.
For typical halo models, this is a function of the radial distance, $r$, strongly peaked towards the Galactic Center.

We consider for the DM distribution the {generalized} Navarro-Frenk-White profile (gNFW) which is described by:
\begin{equation}
\rho_\chi(r) = \frac{\rho_s}{(r/r_s)^\gamma (1+r/r_s)^{3-\gamma}}
,\end{equation}
where $\rho_s$ is a normalization constant and $r_s$ is a characteristic radius below which the profile scales as $r^{-\gamma}$. The value $\gamma = 1$ corresponds to the standard NFW profile~\citep{Navarro1995mnras}.
The profile parameters $\rho_s$ and $r_s$ can be obtained in terms of the virial mass $M_{\rm vir}$ and the concentration parameter $c$.
The virial mass is the mass contained in the virial radius $R_{\rm vir}$. This is defined as the radius of the sphere in which the average DM density is equal to 200 times the critical density of the Universe. The concentration parameter is related to this quantity as $c = R_{\rm vir} / (2-\gamma)r_s$. 

\section{The code characteristics and structure}
\label{sec:code}

\HERMES~is designed with two priorities in mind: ease of use and extensibility. These two priorities dictate the code structure and pattern choices. Our first priority led to the building of an optional self-explanatory Python interface to the code thanks to~\texttt{pybind11}~\citep{pybind11}, by relying on legible functional-style programming as far as modern C++ allows, and by an extensive documentation generated from annotated and commented C++ sources using \texttt{Doxygen}~\citep{doxygen} and \texttt{Sphinx}~\citep{sphinx}. 

Our second priority pushed us to adopt a modular code structure. The modularity means that different logical components of the code, which follow the structure of the previously mentioned physical equations, are separated and independent, and they communicate to each other over standardized and comprehensible interfaces. In addition, a modular code is also easier to read and understand; hence, it contributes to the correctness of the  code. However, the most essential parts ensuring the correctness of the  code are numerous unit tests~\citep{Myers2012} and an improved static type system~\citep{Pierce2002, Gao2017}. The unit tests automatically verify if implemented functions and methods return a correct numerical output for a specified input, while the basic C++ static-type system is enhanced with dimensional analysis of physical quantities and formulas based on~\citet{BartonBook}, so that the dimensional validity of every physical expression is automatically checked by the compiler.

The code draws upon several main designs and toolset choices from \CRPROPA\footnote{\url{https://crpropa.github.io}}~\citep{Batista2016jcap}, a public code for propagating CR particles and inherits magnetic field models, vectors, and grid classes  directly from it. However, as the primary function of \HERMES~significantly differs from~\CRPROPA, the major part of the code is newly developed.
For some pieces, the code follows design choices of \GammaSky, a private code which was successfully exploited to compute gamma-ray models of the Galaxy and is now widely used in experimental collaborations (see, e.g.,~\citealt{Gaggero2015apj,Gaggero2017prl}).

The minimal elements of any computation in \HERMES~consist of a skymap, which is a \HEALPIX\footnote{\url{https://healpix.jpl.nasa.gov}}-compatible~\citep{Healpix2005} container, and an integrator, which performs the LOS calculation for every pixel of the skymap. 
The user defines the properties of the skymap container, its resolution (or the number of pixels), the frequency or energy if needed, and attaches a compatible integrator of a relevant physical process that has to be computed. 
The different integrators require various specific additional components to be specified during the initialization phase, such as for instance a gas density model, a magnetic field model, or a CR flux model. Once all required elements are specified, the computation can be initiated. The final results, contained in the skymap, can be saved, for example, to a FITS-format file. 
The basic \HERMES~workflow is illustrated in Fig.~\ref{fig:flowchart}. 

\tikzstyle{decision} = [diamond, draw, text width=4.5em, text badly centered, node distance=3.5cm, inner sep=0pt]
\tikzstyle{block} = [rectangle, draw, node distance=3.5cm and 1.5cm, text width=5em, text centered, rounded corners, minimum height=3em]
\tikzstyle{rectangular} = [rectangle, draw, text width=5em, text centered, minimum height=3em]
\tikzstyle{line} = [draw, -latex']
\tikzstyle{cloud} = [draw, ellipse, node distance=3.5cm and 1.5cm, text centered, minimum height=2em]

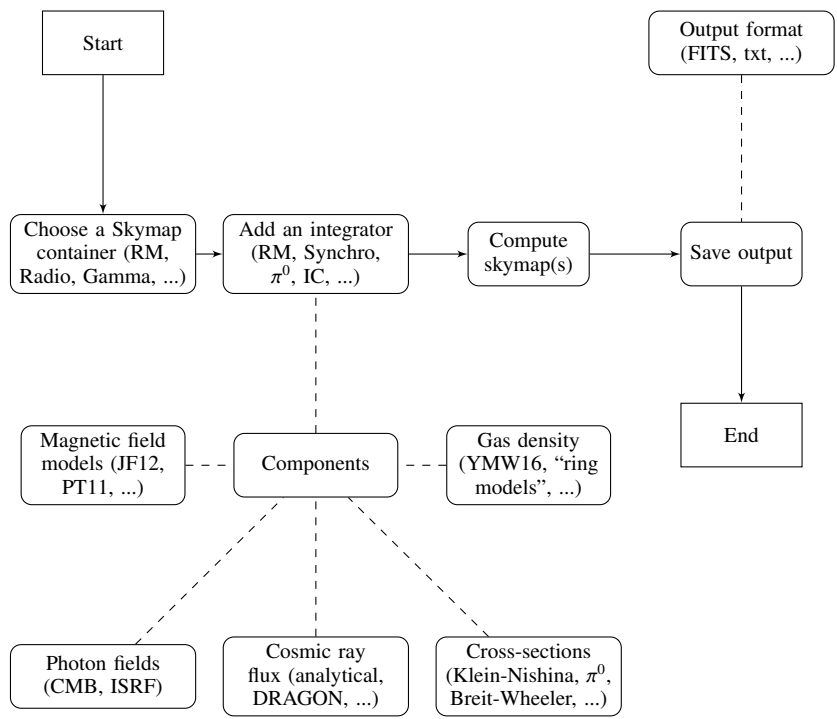
\begin{figure*}
\begin{minipage}[c]{0.66\textwidth}
  \centering
\begin{tikzpicture}[scale=0.8, every node/.style={scale=0.8}, node distance = 3cm and 1.5cm, auto]
    \linespread{1}
    \node [rectangular] (start) {Start};
    \node [block, below of=start, text width=8em] (skymap) {Choose a Skymap container (RM, Radio, Gamma, ...)};
    \node [block, right of=skymap, text width=8em] (integrator) {Add an integrator (RM, Synchro, \(\pi^0\), IC, ...)};
    \node [block, right of=integrator] (compute) {Compute skymap(s)};
    \node [block, right of=compute] (save) {Save output};
    \node [rectangular, below of=save] (end) {End};
    
    \node [block, below of=integrator, text width=7em] (integratormore) {Components};
    \node [block, left of=integratormore, text width=7em] (magnetic) {Magnetic field models (JF12, PT11, ...) };
    \node [block, right of=integratormore, text width=7em] (density) {Gas density (YMW16, ``ring models'', ...) };
    \node [block, below of=integratormore, text width=8em] (photon) {Cosmic ray flux (analytical, DRAGON, ...) };
    \node [block, below of=magnetic, text width=8em] (cr) {Photon fields (CMB, ISRF) };
    \node [block, below of=density, text width=8em] (crossection) {Cross-sections (Klein-Nishina, $\pi^0$, Breit-Wheeler, ...)};
    \node [block, above of=save, text width=8em] (formats) {Output format (FITS, txt, ...)};
    
    \path [line] (start) -- (skymap);
    \path [line] (skymap) -- (integrator);
    \path [line] (integrator) -- (compute);
    \path [line] (compute) -- (save);
    \path [line] (save) -- (end);
    
    \path [draw,dashed] (integratormore) -- (integrator);
    \path [draw,dashed] (magnetic) -- (integratormore);
    \path [draw,dashed] (density) -- (integratormore);
    \path [draw,dashed] (cr) -- (integratormore);
    \path [draw,dashed] (photon) -- (integratormore);
    \path [draw,dashed] (crossection) -- (integratormore);
    
    \path [draw,dashed] (formats) -- (save);
\end{tikzpicture}
\end{minipage}\hfill
\begin{minipage}[c]{0.32\textwidth}
\label{fig:flowchart}
\caption{Flow diagram of basic \HERMES~usage.}
\end{minipage}
\end{figure*}

It is worth emphasizing that the skymap pixels are mutually independent and integrators are generally stateless\footnote{The memory state of integrators is not changed during the computation of a pixel with the exception of the caching mechanism which can be enabled for certain integrators to gain performance in some cases.}, and so the computation process is thread-safe, and therefore the total computation time in general decreases {linearly} with the number of threads available. 
The parallelization is implemented using the native C++11 thread library to ensure cross-platform compatibility.

\subsection{Skymap classes}

All skymap classes are initialized first by providing the \HEALPIX~\texttt{nside} parameter\footnote{The \HEALPIX~terminology names three related map parameters: the resolution ({\tt Res}), the number of pixels per side ({\tt NSide}), and the total pixel number ({\tt NPixels}). 
These relate as follows: $\mathtt{NSide} = 2^\mathtt{Res}$ and $\mathtt{NPixels} = 12~\mathtt{NSide}^2$.} which determines the total pixel number of a map. A larger number of pixels leads to a higher angular resolution of the skymap.

Several types of skymap classes are available, and they are distinguished based on the physical units of pixels. For example, the rotation measure skymap, \texttt{RotationMeasureSkymap}, stores pixels with the \([\mathrm{rad}\,\mathrm{m}^{-2}]\) dimension, the radio skymap defined for a given frequency, \texttt{RadioSkymap}, stores temperature pixels in Kelvins, while the gamma skymap for a given energy, \texttt{GammaSkymap}, stores the differential intensity pixels. Furthermore, for the radio and gamma-ray skymaps, a stacked container abstraction is provided. This stacked container can store multiple skymaps of the same type, ranging in frequency or energy; these are named \texttt{RadioSkymapRange} and \texttt{GammaSkymapRange}.

To each type of skymap, one can attach a specific integrator, map masks, and an output format, either using the class setters or constructors. Integrators are presented in the following section. The map masks are sets of rules specified in the Galactic coordinates and exclude certain regions of the sky from being computed. These so-called unseen pixels are set to \(-1.6375 \times 10^{30}\), following the \HEALPIX~convention. 
For example, the \texttt{RectangularWindow} class, for two given sets of Galactic latitudes and longitudes, excludes from the computation everything outside of the rectangular window that is stretched between those two coordinate sets. 
Other provided mask classes, named according to their purpose, are \texttt{CircularWindow}, \texttt{MaskList}, and \texttt{InvertMask} where the last two take mask classes as their arguments;  for example, one can  use \texttt{MaskList}
 to combine several rectangular and circular windows into one mask.
Finally, by specifying the output format, one determines how the calculated skymap will be saved after calling its \texttt{save} method. Although the FITS format is provided by default and is generally recommended because it is supported by other software packages such as \texttt{healpy}~\citep{Zonca2019}, other save mechanisms can be implemented easily by inheriting the \texttt{Output} abstract class.

\subsection{Integrator classes}

Skymaps only provide the storing facilities for pixels, while integrators evaluate pixels for the directions specified in the traditional spherical coordinates~\citep{iso200980000}. 
Each integrator performs the line of sight integration according to the appropriate equations described in sect.~\ref{sec:physics}, and returns the result in the form of a quantity matching the related skymap class quantity to which the integrator is attached. The integrator classes implemented in \HERMES~are as follows.
\begin{itemize}
    \item {\tt DispersionMeasureIntegrator} is an integrator used in conjunction with \texttt{DispersionMeasureSkymap} and takes a model for the distribution of free (thermal) electrons as an input and returns the dispersion measure in [pc~cm$^{-3}$].
    
    \item {\tt RotationMeasureIntegrator} requires two components: a Galactic field model and a free electron density model following Eq.~\ref{eq:rotationmeasure}. It returns the rotation measure in [rad~m$^{-2}$].
    
    \item {\tt SynchroIntegrator} takes the magnetic field model and the distribution of CR leptons as input, and returns the synchrotron intensity.
    
    \item {\tt SynchroAbsorptionIntegrator} takes the magnetic field model, the distribution of CR leptons, and the distribution of free electrons as input, and returns the synchrotron intensity, with the free-free absorption taken into account.
    
    \item {\tt FreeFreeIntegrator} takes a model of the distribution of free (thermal) electrons as an input and returns the free--free emissivity (method {\tt spectralEmissivity}) and absorption coefficient (method {\tt absorptionCoefficient}).
    
    \item {\tt PiZeroIntegrator} requires as inputs the (hadronic) CR distribution, the (neutral) gas model, and the cross-section model, and computes the intensity of the $\pi^0$ emission (by summing over the gas rings).
    
    \item {\tt PiZeroAbsorptionIntegrator} takes the ISRF model as additional input with respect to the ones taken by {\tt PiZeroIntegrator}. The output is the attenuated intensity of the $\pi^0$ emission.    
    
    \item {\tt BremsstrahlungIntegrator} requires as inputs the (leptonic) cosmic-ray distribution, the (neutral) gas model, and the cross-section model, and computes the bremsstrahlung intensity.
    
    \item {\tt InverseComptonIntegrator} receives as inputs the (leptonic) cosmic-ray distribution, the ISRF model, and the cross-section model, and computes the IC intensity by integrating the emissivity along the LOS.
    
    \item {\tt DarkMatterIntegrator} requires the profile of the Galactic DM halo, the (prompt) gamma-ray spectrum (normalized to one annihilation), and computes the integrated prompt gamma-ray emission.
\end{itemize}

The examples presented in Section 4 make use of these integrators.

\subsection{Astrophysical components as integrator inputs\label{sec:components}}

Astrophysical fields, particle distribution profiles, interaction cross-sections, and other physical models serve as inputs for the above-described integrators; they define the physical content of any skymap calculation in \HERMES.

The code modularity encourages users to modify available components or to adapt and implement new ones. The only requirement for these is to inherit appropriate abstract classes that the given integrator accepts and to implement the corresponding \texttt{get} method(s). In principle, for each look-up, a component can either analytically calculate the required value or extract the value from pre-computed data saved in tables or multidimensional grids where the value in between grid points is interpolated. Both types of look-ups can be delegated to external libraries. 

Various tabulation and Cartesian-based grid classes and utilities, such as interpolation techniques or reading and loading functions, are supplied by default to ease the implementations of new components.

In Section~\ref{sec:applications} we introduce several components that we use to present example implementations of the code.
Most of these components are either adopted from publicly available codes licensed under the compatible copyleft licences, such as~\texttt{YMW16} and~\texttt{JF12}, or newly implemented following original papers.
In addition, the models \texttt{Kamae06\{Gamma,Neutrino\}} depend on an external library, \texttt{cparamlib}\footnote{The source code of the library with routines written in C is publicly available at \url{https://github.com/niklask/cparamlib}.}, shipped with the code.

\subsection{Code usage}

An important step has been made in designing \HERMES~following modern standards for modular codes. 
Different aspects of the simulation (e.g., magnetic fields, CR density,~etc.) are separated into modules. 
Each module is independent from other modules, and therefore each module can be replaced or a new module can be added, making \HERMES~a flexible framework that can be extended without the need to modify other components. 
The only requirement for an additional user-provided module is to respect the general module C++ interface, as already explained in Sect. \ref{sec:components} for astrophysical components, but applies also for any other module, such as integrators, skymaps, output modules, and skymap masks.

In a user-defined simulation, a sequence of these independent modules is introduced, combined, and wired up. This can be accomplished either through a new C++ program compiled and linked against the \HERMES~headers and library or via the Python interface imported as \texttt{pyhermes}, a pre-compiled Python module.

A series of examples written in C++, Python, and as Jupiter notebooks illustrate how to use the specific modules in order to compute the maps or the spectra. These are available in a separate repository\footnote{\url{http://github.com/cosmicrays/hermes-examples}}.
Here we present one of the examples provided to compute the $\pi^0$ map in order to illustrate the Python interface in more detail. 

The example starts with the following initialization commands and general settings:
\begin{lstlisting}[language=Python]
from pyhermes import *
from pyhermes.units import *

nside = 512
Egamma = 0.1*TeV
obs_pos = Vector3QLength(8.0*kpc,0*pc,0*pc)
\end{lstlisting}

Here we have set a \HEALPIX ~resolution of 512  (corresponding to $\simeq 3.1 \cdot 10^6$ pixels with mean spacing of $\simeq 0.1^\circ$), a reference energy of $100$ GeV, and the coordinates associated to the observer (Sun) position in a standard Cartesian grid.

We then specify the astrophysical components:
\begin{lstlisting}[language=Python, firstnumber=7]
dragon2D_proton =\
    cosmicrays.Dragon2D(Proton)
dragon2D_helium =\
    cosmicrays.Dragon2D(Helium)
cr_list = [dragon2D_proton,
           dragon2D_helium]
kamae_crosssection =\
    interactions.Kamae06Gamma()
neutral_gas_HI =\
    neutralgas.RingModel(
        neutralgas.GasType.HI)
\end{lstlisting}

In this block, we have instructed \HERMES~to load the key components of the computation (see Fig.~\ref{fig:flowchart} for a schematic diagram of the different classes involved). These default ingredients are further described in the following section.

In particular we have set:
\begin{itemize}
    \item the spatial distribution of CR hydrogen and helium nuclei taken (in this case) from the output of a reference \DRAGON\footnote{\url{https://github.com/cosmicrays/}}~run (typically provided in {\tt FITS} format);
    \item the cross-section model from~\cite{Kamae2006apj};
    \item a reference model for the neutral gas distribution (see the following section for more details).
\end{itemize}

Let us now construct and initialize the appropriate integrator class for the $\pi^0$ emission, passing as input parameters the components defined above:
\begin{lstlisting}[language=Python, firstnumber=18]
integrator = PiZeroIntegrator(cr_list,\
        neutral_gas_HI, kamae_crosssection)
integrator.setObsPosition(obs_pos)
\end{lstlisting}
The next integrator command is optional and serves to speed up the calculation. It is discussed in detail in the following section.
\begin{lstlisting}[language=Python, firstnumber=21]
integrator.setupCacheTable(100, 100, 20)
\end{lstlisting}
We proceed with the construction of a new skymap class:
\begin{lstlisting}[language=Python, firstnumber=22]
skymap = GammaSkymap(nside=nside,
                     Egamma=Egamma)
top_left_edge = [5*deg, 20*deg]
bottom_right_edge = [-5*deg, 60*deg]
mask = RectangularWindow(\
          top_left_edge, bottom_right_edge)
skymap.setMask(mask)
skymap.setIntegrator(integrator)
\end{lstlisting}
In this block we have instantiated the {\tt GammaSkymap} class, and attached the already prepared {\tt PiZeroIntegrator} to it. We have also set a mask associated to a region of interest:  a portion of the Galactic plane ($20^\circ < l < 60^\circ$; $-5^\circ < b < 5^\circ$).

Once all the relevant classes are instantiated and initialized, we call the method that contains the core of the computation:

\begin{lstlisting}[language=Python, firstnumber=30]
skymap.compute()
\end{lstlisting}

We are now ready to store the result in a {\tt numpy} array (containing the map in \HEALPIX ~format computed at $100$ GeV), and visualize the map:

\begin{lstlisting}[language=Python, firstnumber=31]
import healpy

hermes_map = np.array(skymap)
healpy.mollview(hermes_map)
\end{lstlisting}

\subsection{Code performance}

\HERMES~is predominantly a CPU-intensive code because of several nested loops originating from  integration procedures listed in Sect. \ref{sec:physics}. But before discussing the computation time, let us present an overview of the working memory  requirements which come, in general, from storing skymap pixels and loading model data. 
For example, 3.1M double precision floating-point pixels are needed to store one \(\mathrm{nside}=512\) map.
For ranges of skymaps, such as \texttt{RadioSkymapRange} or \texttt{GammaSkymapRange}, this size is multiplied by the number of stacked maps in the range. Consequently, the total consumption of skymap containers is of the order of hundreds of megabytes, depending also on the computer platform used. In comparison, loading some astrophysical components 
to memory can require considerably more if dense grids are employed, such as in \texttt{RingModel} (\(\sim\) 800 MB). The requirements on disk storage of the skymaps and model data is reduced compared to the working memory due to file compression.

The central point of the code performance considerations is the computation time of a \HERMES-based program. The computation time is highly sensitive to the execution of the innermost integrand during the LOS integration, especially when triple integrals are used, such as in inverse Compton, Eq. \ref{eq:ICemisssivity}, or pion decay, Eq. \ref{eq:pi0emissivity}, or quadruple integrals, such as scenarios with the absorption. \HERMES~automatically employs the multi-threading feature to distribute the LOS integral computation of pixels over all available CPU threads, consequently halving the computation time with respect to the number of threads. The number of threads employed can be controlled with the \texttt{HERMES\_NUM\_THREADS} environment variable; for example, \texttt{HERMES\_NUM\_THREADS=1} will execute the program in a single thread.

Moreover, the computation time can also be reduced with a caching feature (sometimes called memoization), in which the CPU time is traded for the memory consumption. For example, one can pre-compute the innermost integrand in Eq. \ref{eq:ICemisssivity} or Eq. \ref{eq:pi0emissivity} and store the values in a spatial (3D) grid. Then, during the LOS integration, the procedure only accesses and interpolates the pre-computed grid values instead of calculating the integral over and over again. However, in some cases the repeated calculation of the innermost integrand can be faster than the grid look-up and interpolation calls, especially in the multi-threading context where many threads are slowed down by accessing the same shared memory, making the caching feature inefficient.

In \HERMES, the cache is calculated before the LOS computation and turned on using the \texttt{setupCacheTable(\(N_X\), \(N_Y\), \(N_Z\))} method in gamma-ray-related integrators where \(N_X\), \(N_Y\), \(N_Z\) are the number of grid points for a galaxy of fixed size
\(L_x, L_y = 60~ \mathrm{kpc}\) and \(L_z = 10~\mathrm{kpc}\) with the GC in the middle of the grid. This feature will be expanded in future versions of \HERMES~to allow further customization. For optimal performance, the density of the caching grid should match the density of the sparsest grid model used in the innermost integral calculation, such as CR density.

For the example given in the previous section (\(\mathrm{nside} = 512\)), the measured simulation run time\footnote{Measured with the \texttt{hyperfine} command-line benchmarking tool on 20 cores / 40 threads @ 2.2GHz; HERMES was compiled with GCC 10.2.1.} is (30.1 $\pm$  0.3)~s. The same example, but without caching enabled, that is, with line 21 commented out, 
the measurement gave (1144 $\pm$ 3)~s, or approximately 40 times more than with caching. On the other hand, without any skymap mask applied, that is, for the full sky coverage and with caching, the same example is evaluated in (315 $\pm$ 5)~s.

An average run time per skymap pixel, \(t_\mathrm{pixel}\), can be a useful unit with which to estimate the total run time of a simulation simply by evaluating \(t_\mathrm{total} = N_\mathrm{total} / N_\mathrm{threads} \times t_\mathrm{pixel}\) where \(N_\mathrm{total}\) is the total number of unmasked pixels which should be computed and \(N_\mathrm{threads}\) is the number of system threads available. For available processes in realistic scenarios with caching enabled, average run-times per pixel are given in Table \ref{tab:pixeltimes}. The pixel run time of each process significantly depends on the integrator components attached, and not only the integrator implementation, and therefore the code run time can be reduced by optimizing component value retrieval. The results in Table \ref{tab:pixeltimes} can vary depending on the computer system; they are simply a rough estimate of simulation run times on other systems.

\begin{table}[]
\centering
\caption{Average pixel run times for available processes}
\label{tab:pixeltimes}
\begin{tabular}{|c|l|r|}
\hline
\# & \textbf{Physical process} & \(t_\mathrm{pixel} / \mathrm{ms}\) \\ \hline \hline
1. & Dispersion measure & 10.3 \\ \hline
2. & Rotation measure & 13.7 \\ \hline
3. & Synchrotron & 22.7 \\ \hline
4. & Free-free & 13.9 \\ \hline
5. & Pion decay & 5.4 \\ \hline
6. & Bremsstrahlung & 7.1 \\ \hline
7. & Inverse Compton & 117.5 \\ \hline
8. & Dark matter & 0.3 \\ \hline
\end{tabular}
\end{table}

\section{Some relevant applications}
\label{sec:applications}

In order to demonstrate the capabilities of \HERMES, here we present a set of simulated diffuse emission skymaps and spectra computed using up-to-date models of the relevant Galactic components.
The aim of this section is to demonstrate the capability of \HERMES~and, at the same time, the potential of a multi-wavelength approach, linking radio and gamma-ray data self-consistently to infer the properties of the CR population on Galactic scales.

In the following sections, we describe the astrophysical ingredients implemented in a reference setup for \HERMES, we provide a comprehensive visualization of the skymaps and spectra at different wavelengths, and we comment on the main features of the resulting emission templates associated to the physical processes discussed above.
All input models presented here are optional and can be easily substituted by other models.

For the CR distributions of leptons and nuclei, we adopt the CR densities obtained by solving the CR diffusion-loss equation (including re-acceleration) under the assumption of a homogeneous and  isotropic diffusion coefficient by means of the \DRAGON~code~\citep{Evoli2017jcap,Evoli2018jcap}. 
All the assumptions and parameters of the adopted CR model are discussed in~\cite{Fornieri2020jcap}, where the authors compared their predictions with the most relevant local CR observables (namely: proton, helium, carbon and oxygen flux, boron-to-carbon ratio, low-energy lepton and antiproton fluxes) over a wide energy range (from $\simeq 10$ MeV up to $\simeq 1$ TeV).

The map calculations are performed using NSIDE=256, corresponding to an angular resolution of $\sim 0.2^\circ$.

\subsection{Faraday rotation and dispersion measures of the Galactic magnetic fields}
\label{sec:faraday}

\begin{figure}[t]
\centering
\includegraphics[width=0.93\columnwidth]{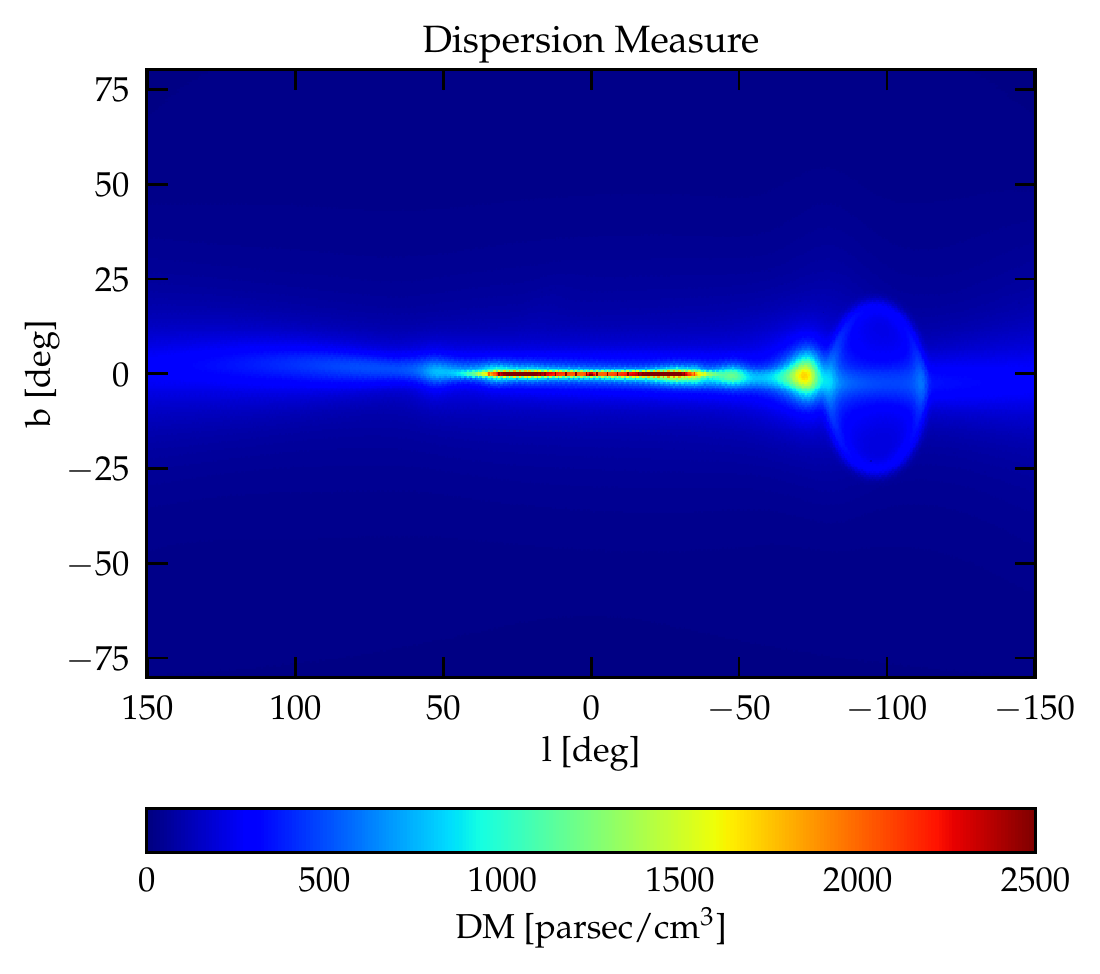}
\caption{Cartesian projection in Galactic coordinates of the dispersion measure computed with the free electron model by~\cite{Yao2017apj}.}
\label{fig:dispersionmeasure-skymap}
\end{figure}

\begin{figure*}
\centering
\includegraphics[width=0.93\columnwidth]{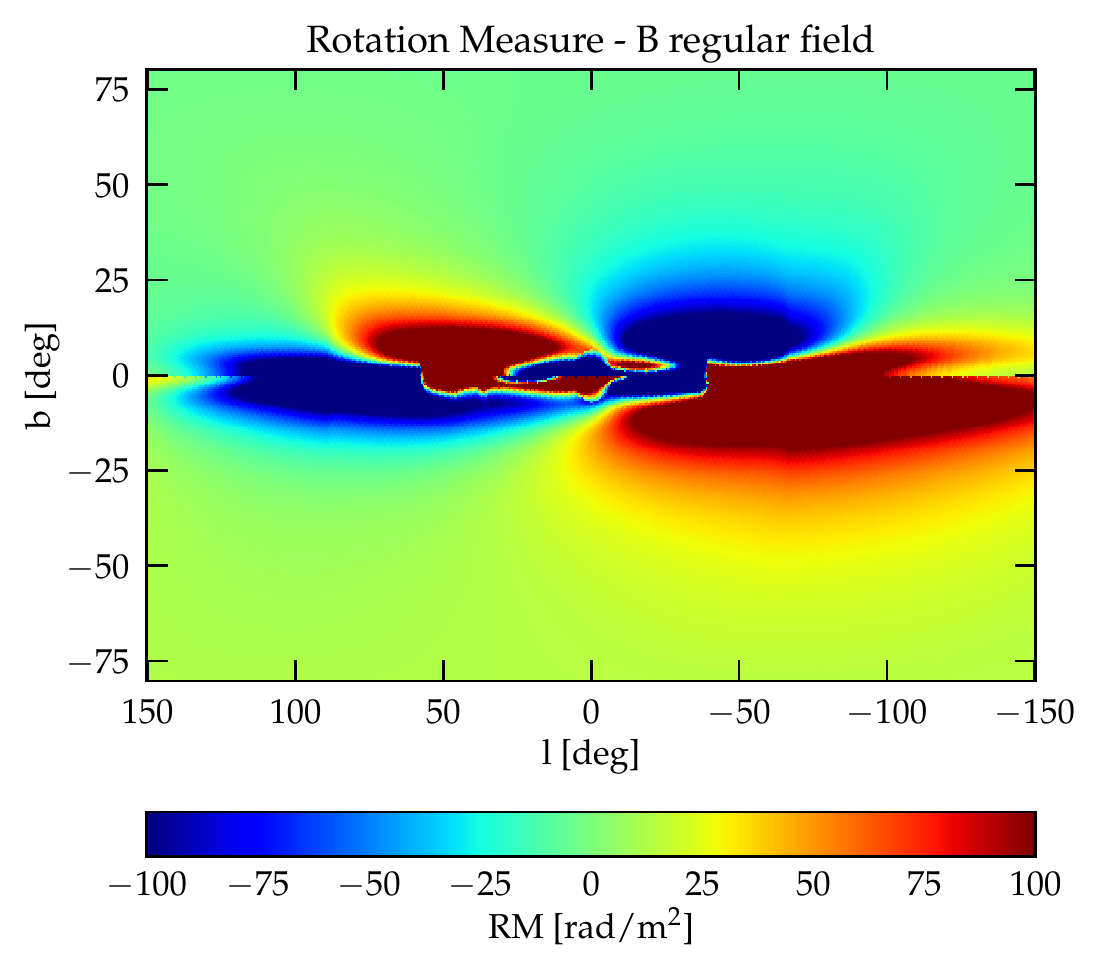}
\includegraphics[width=0.93\columnwidth]{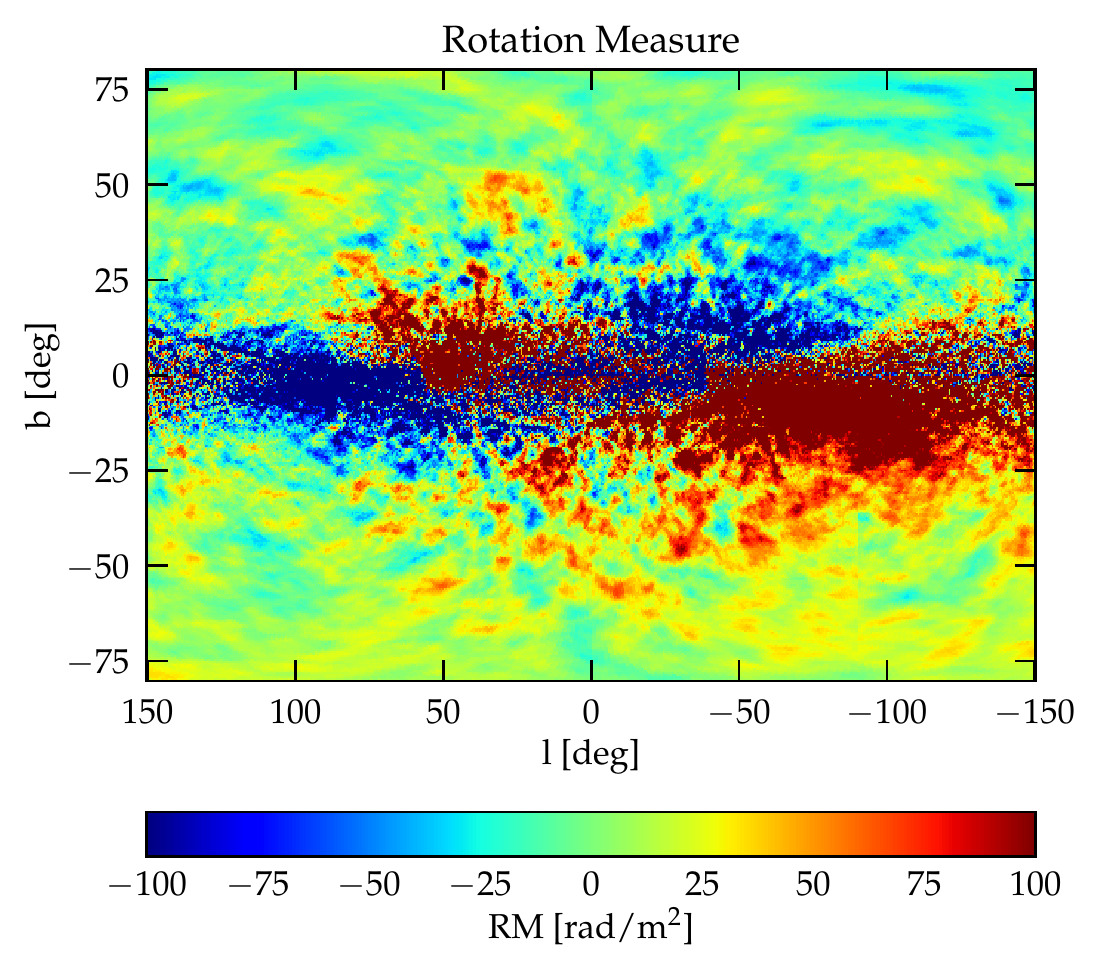}
\caption{Cartesian projection in Galactic coordinates of the rotation measure. {\it Left panel:} Map computed with the large-scale Galactic magnetic field model provided by~\cite{Jansson2012apj}. {\it Right panel:} Computation including a model of the turbulent field computed as detailed in Sect.~\ref{sec:faraday}.}
\label{fig:rotationnmeasure-skymap-noturb}
\end{figure*}

The full-sky map of Faraday rotation and dispersion measure can be used to derive the properties of the large-scale structure of the Galactic magnetic field and of the electron density.

To simulate these maps, we considered the three-dimensional model by~\cite{Yao2017apj}  for the free electrons in the Galaxy.  
The model features an extended thick disk that represents the so-called warm interstellar medium (WIM), a thin disk representing the Galactic molecular ring,  a spiral arm pattern (based on a recent fit to Galactic HII regions), a GC disk, and several local features (including the Gum Nebula, Galactic Loop I, and the Local Bubble). 
The parameters of this model are fit on a wide set of distance measures and distances of Galactic pulsars.
        
Concerning the large-scale Galactic magnetic field (GMF) we adopt the model by~\cite{Jansson2012apj}, with the updated parameters of~\cite{Unger2017icrc}\footnote{Available at~\url{https://github.com/CRPropa/CRPropa3/blob/3.1.6/src/magneticField/JF12Field.cpp}}. 
The model entails a disk component (which follows a spiral-arm pattern), a toroidal halo component, and an axisymmetric and poloidal out-of-plane component. 
The model is based on a best fit of a comprehensive set of data, including the WMAP7 maps of synchrotron emission and rotation measures of $4 \times 10^4$ extragalactic sources.

The turbulent magnetic field model is computed by a class obtained from \CRPROPA~\citep{Batista2016jcap}, where the field is implemented following an approach firstly developed by~\cite{Giacalone1994apj}.
%
The field is initially constructed in a discrete $k$-space \(\mathbf{B}(\mathbf{k})\) by drawing for each grid point (in total N$^3$) a randomly oriented vector \(\mathbf{k}\) which satisfies solenoidality, i.e., \(\mathbf{k}\cdot\mathbf{B} = 0\).
The vector amplitude is determined from a turbulent power spectrum \(\mathbf{B}^2(k) = k^{\alpha}\) where the default value \(\alpha = -11/3\) represents the Kolmogorov spectrum in 3D space. The amplitude is further modulated by a random complex phase.   
Every \(\mathbf{B(\mathbf{k})}\) that lies outside of the range \(\left[ k_{\rm min} \equiv \mathrm{spacing} / L_{\rm max}, k_{\rm max} \equiv \mathrm{spacing} / L_{\rm min}\right]\) is set to zero, where \(k_{\rm min}\) and \(k_{\rm max}\) are the minimum and maximum wavenumbers and \texttt{spacing} is the physical distance between two nearest grid points. In doing so, \(L_\mathrm{max}\) and \(L_\mathrm{min}\) are the maximal and minimal physical scale of turbulence and are user-defined quantities.
Subsequently, the k-field is transformed into real space with a Fast Fourier Transform method provided from the \texttt{FFTW} software package~\citep{1386650}. Finally, the grid is normalized at each position to the local value of \(B_{\rm rms}\) given by the magnetic field model of~\cite{Jansson2012apj}. For  details of the turbulent field implementation see, for example, Appendix B in \cite{dundovic2018anisotropies}.

 Figure~\ref{fig:dispersionmeasure-skymap} shows our full-sky map of the dispersion measure in which the contribution of the Galactic plane is visible, as are several local structures associated with the adopted electron density distribution.
In Fig.~\ref{fig:rotationnmeasure-skymap-noturb} we visualize the rotation measure sky map and we compare the case with and without the turbulent magnetic field.

\subsection{Radio emissions}

\begin{figure*}
\centering
\includegraphics[width=0.93\columnwidth]{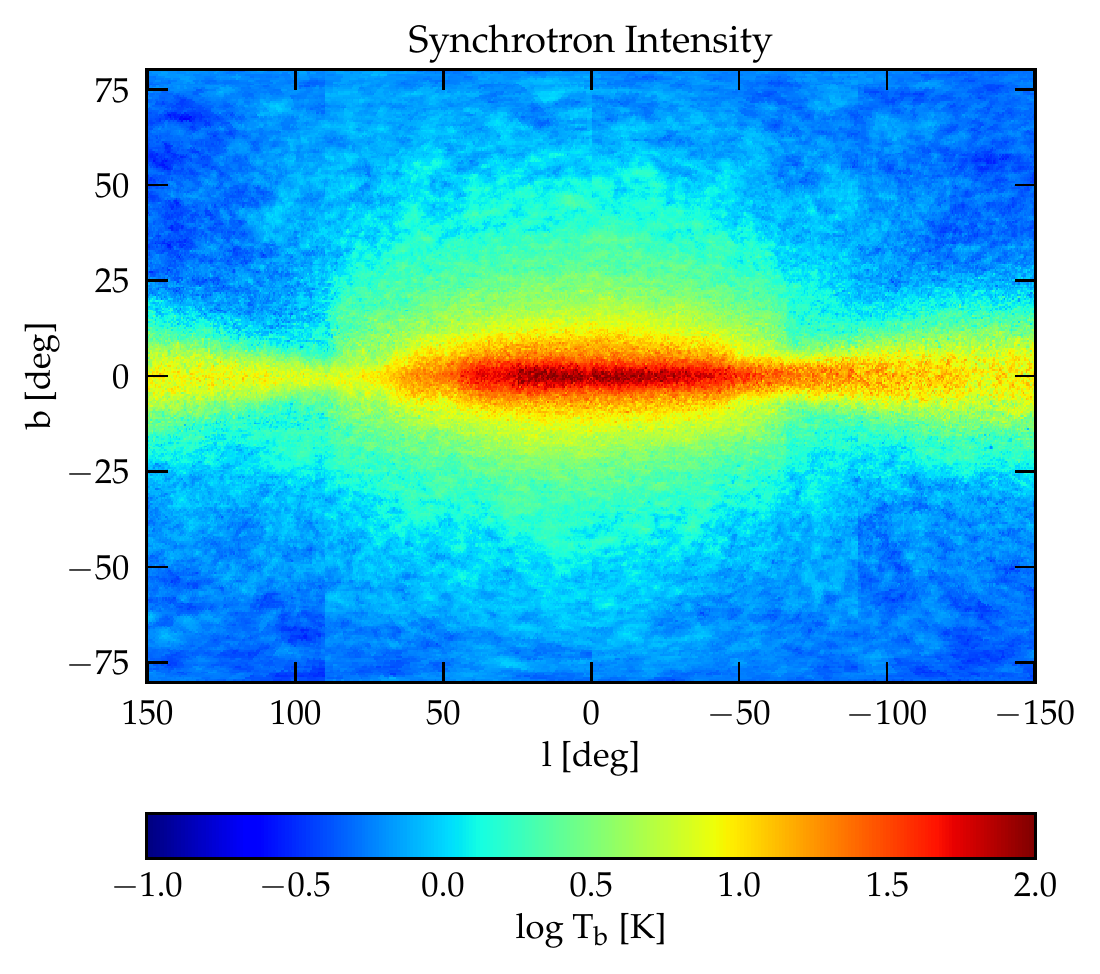}
\includegraphics[width=0.93\columnwidth]{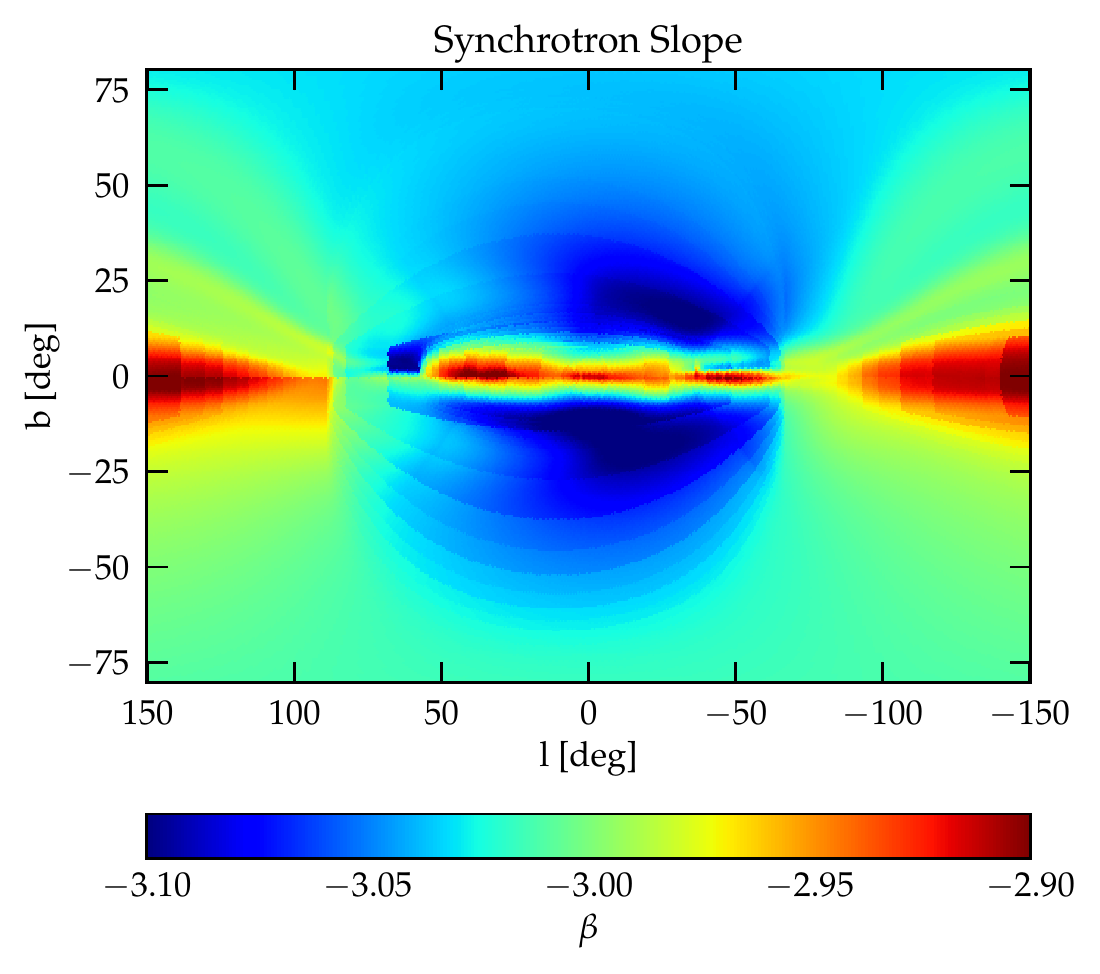}
\centering
\caption{Cartesian projection in Galactic coordinates of the synchrotron intensity ({\it left panel}) and of the synchrotron slope ({\it right panel}). Both plots are computed at the frequency of 408 Mhz.}
\label{fig:synch-skymap}
\end{figure*}

\begin{figure*}
\centering
\includegraphics[width=0.93\columnwidth]{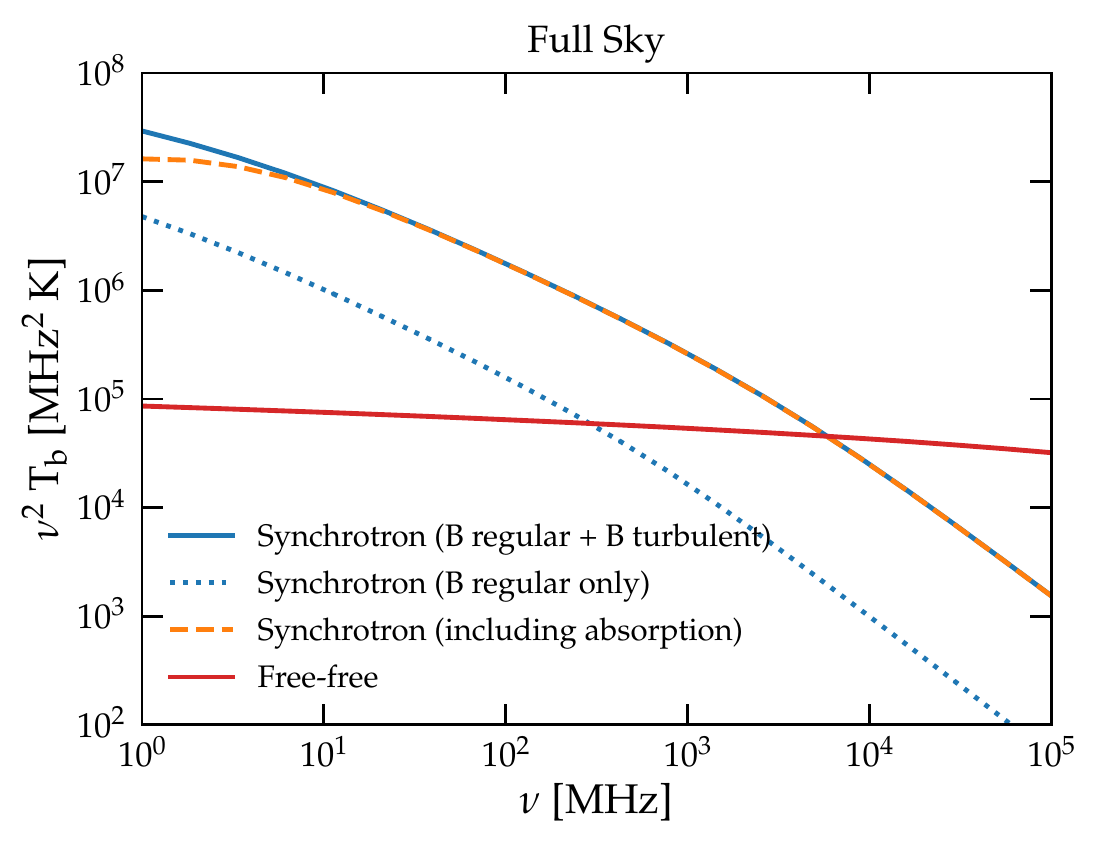}
\includegraphics[width=0.93\columnwidth]{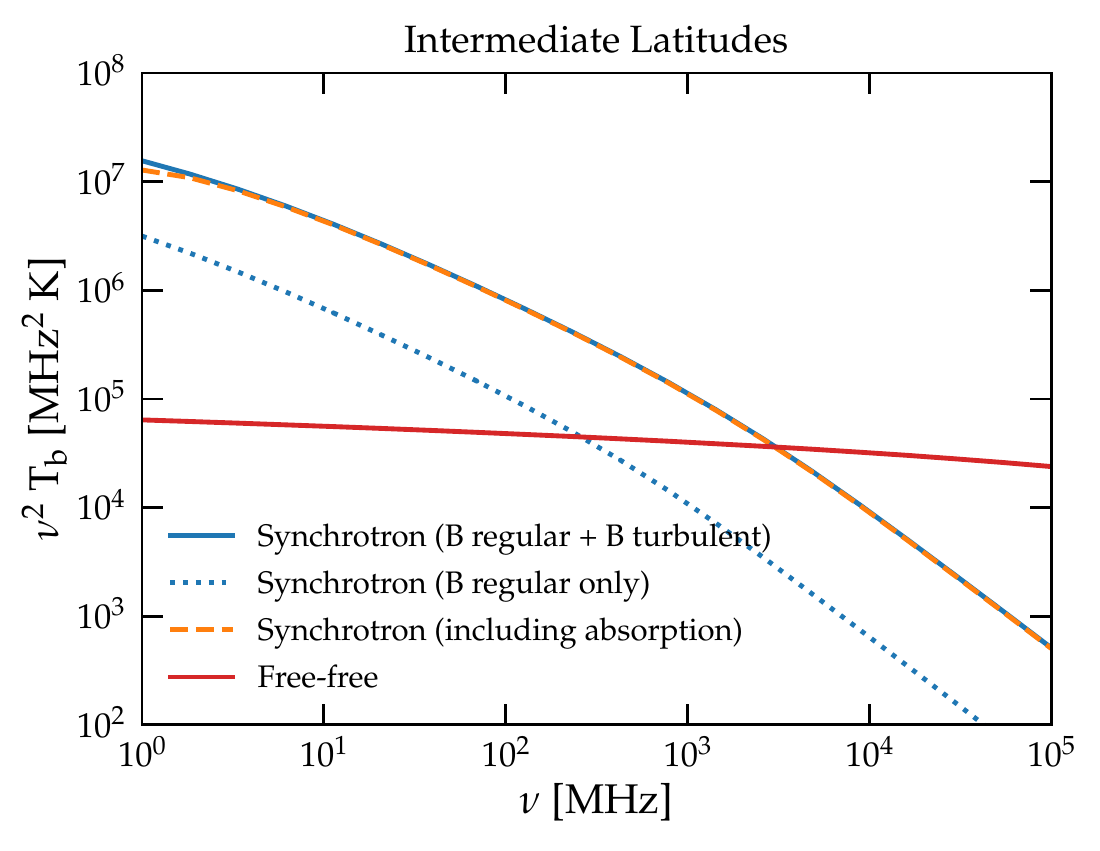}
\caption{Integrated spectra associated to the most relevant processes in the radio domain: synchrotron emission (without or with free--free absorption) and free--free emission. {\it Left panel:} Full-sky integration. {\it Right panel:} Intermediate-latitude region defined by  $10^{\circ} < b < 40^{\circ}$ }
\label{fig:radio-spectra}
\end{figure*}

 Figure~\ref{fig:synch-skymap} (panel A) shows our the sky map of the simulated synchrotron intensity.
We compute the map at the reference frequency of $408$ MHz, because it is widely covered by high-quality all-sky continuum surveys (see for instance~\citealt{Haslam1981aa}), and the Galactic synchrotron emission is the dominant contribution.
The dominant contribution from the Galactic plane can be clearly identified in the map. 

The spatial variation of the spectral slope at that reference frequency is also shown in Fig.~\ref{fig:synch-skymap} (panel B). 
This slope is directly connected to the slope of the population of CR electrons at a few GeV. A clear trend is the progressive softening towards greater latitudes, as expected when the portion of the leptonic CR population that is probed is located farther away from the regions where most of the acceleration takes place.

The synchrotron spectrum in the sky region (for both full sky and the intermediate-latitude region defined by  $10^{\circ} < b < 40^{\circ}$) is compared in Fig.~\ref{fig:radio-spectra} with the spectrum of free--free emission.
We notice in particular that synchrotron emission is the dominant process up to $\simeq 10$ GHz, the band at which free--free emission, featuring a harder spectrum, starts to take over. 
The impact of absorption is small. Only at small frequencies (around 1 MHz) is the effect around 10-20\%.

\subsection{Gamma-ray sky below 1 TeV}

\begin{figure*}
\centering
\includegraphics[width=0.93\columnwidth]{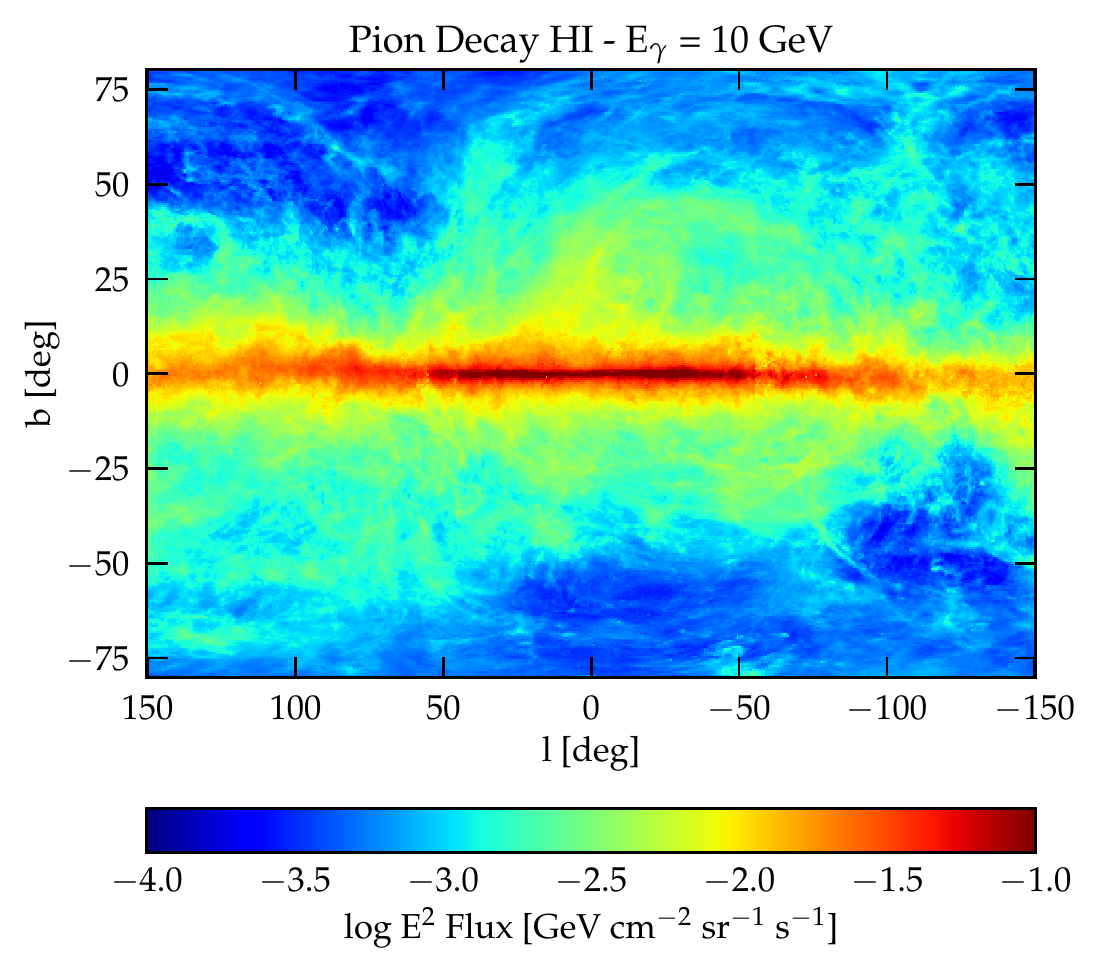}
\includegraphics[width=0.93\columnwidth]{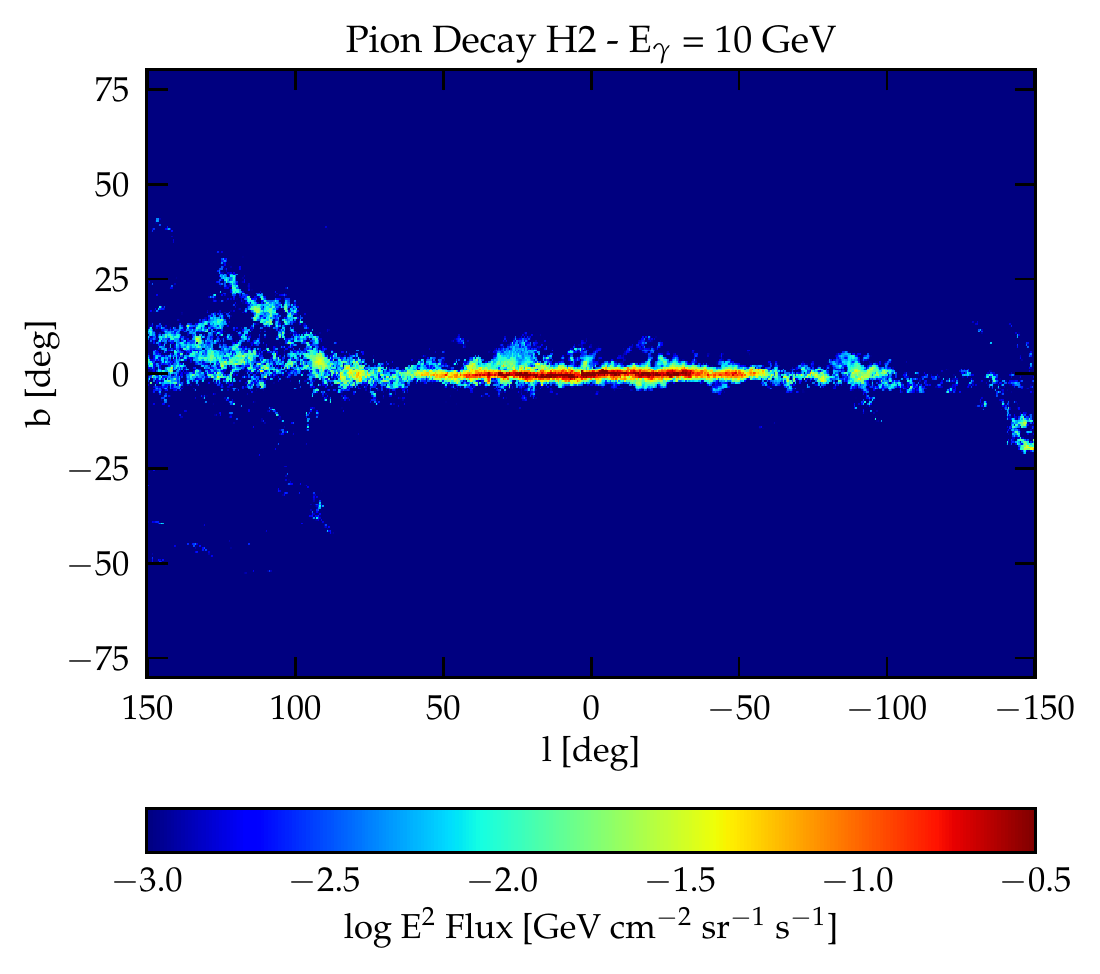}
\caption{Gamma-ray sky map at 10 GeV. We show the Cartesian projection in Galactic coordinates of the $\pi^0$-decay gamma-ray flux at $E_\gamma = 10$~GeV. {\it Left panel:} Gamma-ray emission associated to $\pi^0$ production on HI. {\it Right panel:} Gamma-ray emission associated to $\pi^0$ production on $H_2$.}
\label{fig:gammaray-skymap-pi0-HI-10GeV}
\end{figure*}

\begin{figure}
\centering
\includegraphics[width=0.93\columnwidth]{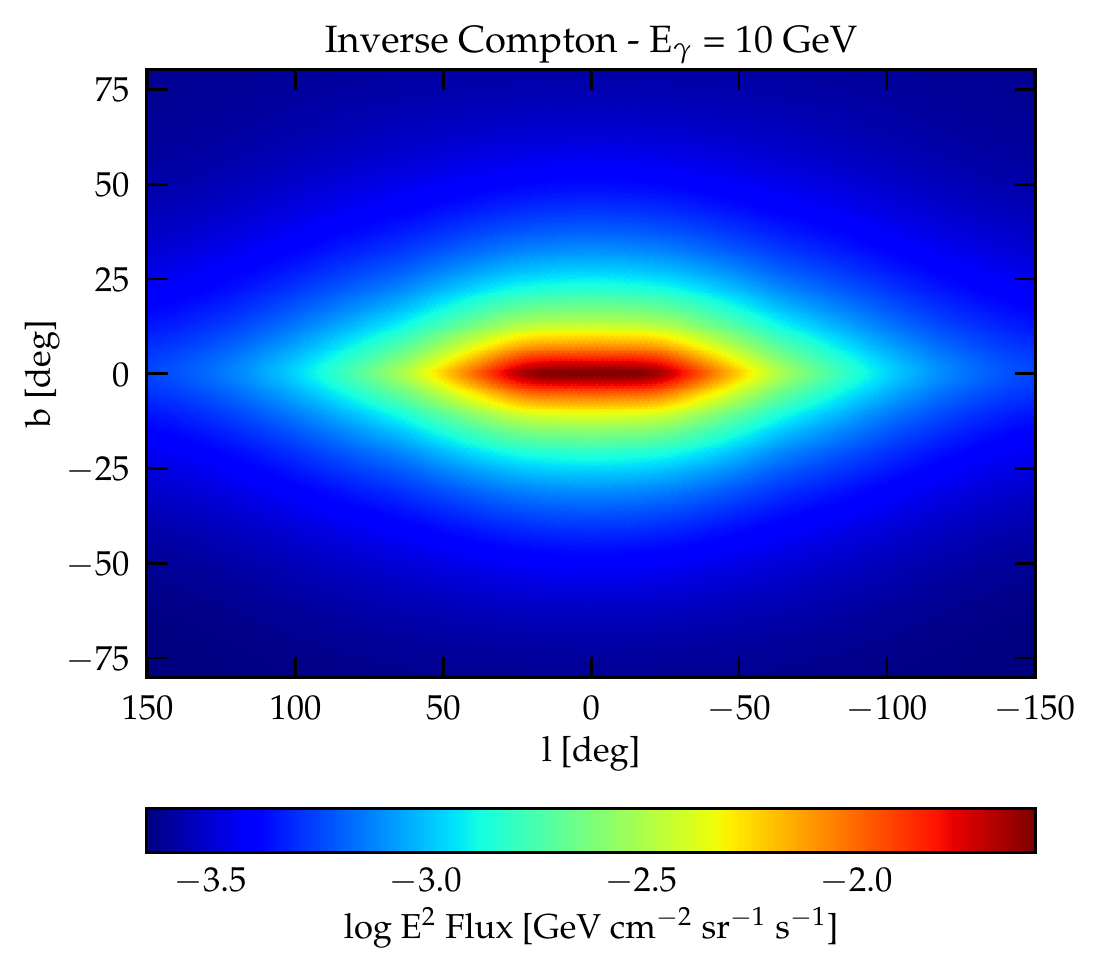}
\caption{Cartesian projection in Galactic coordinates of the IC gamma-ray flux at $E_\gamma = 10$~GeV.}
\label{fig:gammaray-skymap-IC-10GeV}
\end{figure}

\begin{figure*}
\centering
\includegraphics[width=0.9\columnwidth]{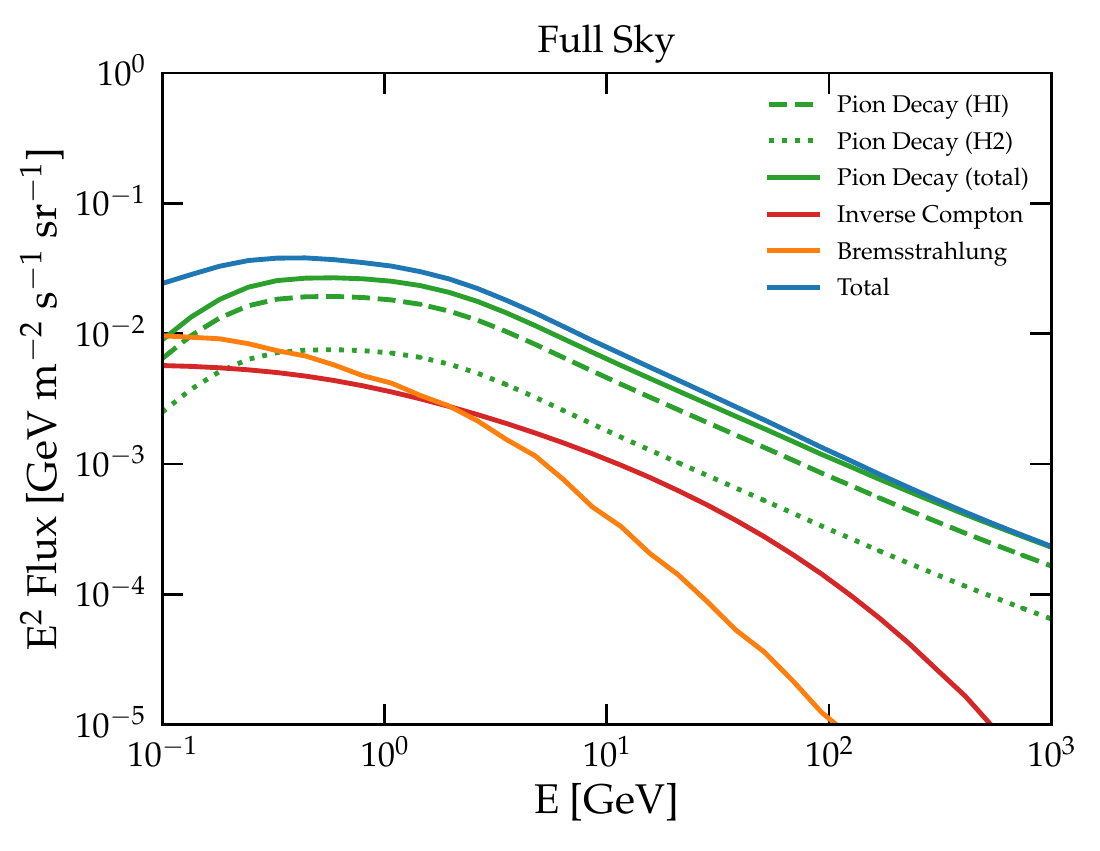}
\includegraphics[width=0.9\columnwidth]{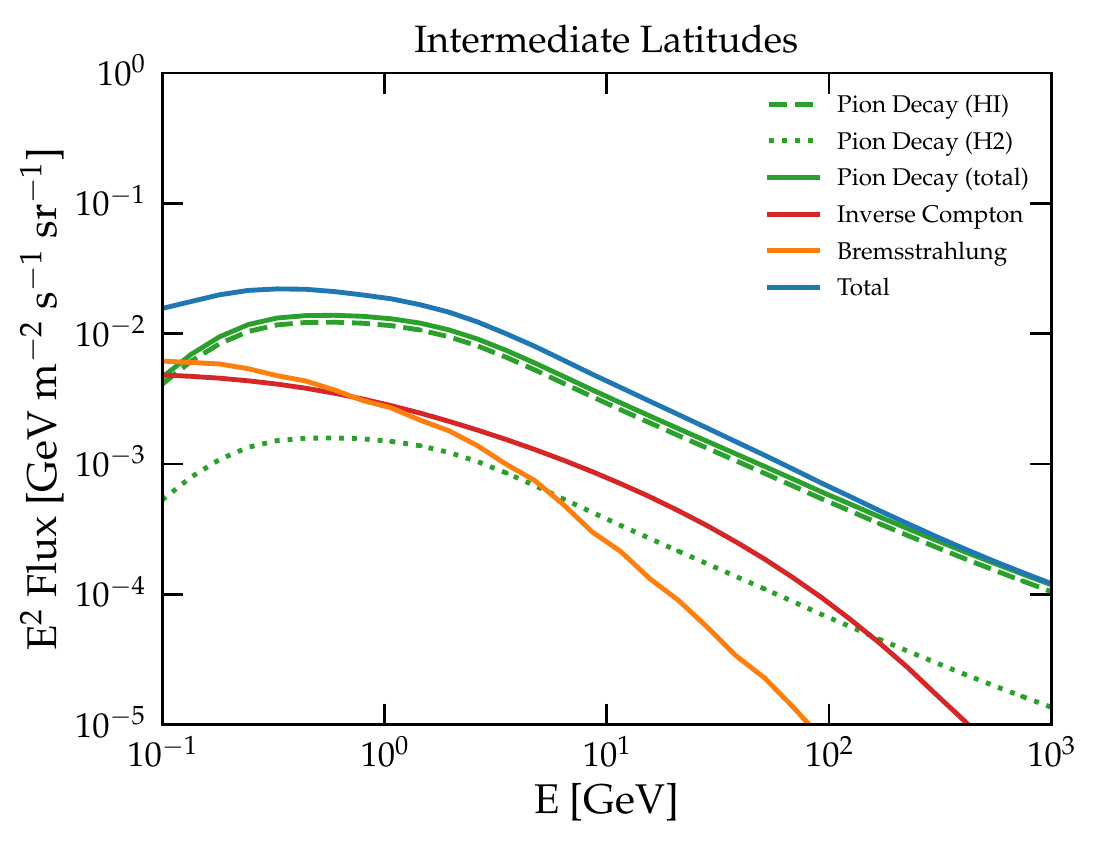}
\includegraphics[width=0.9\columnwidth]{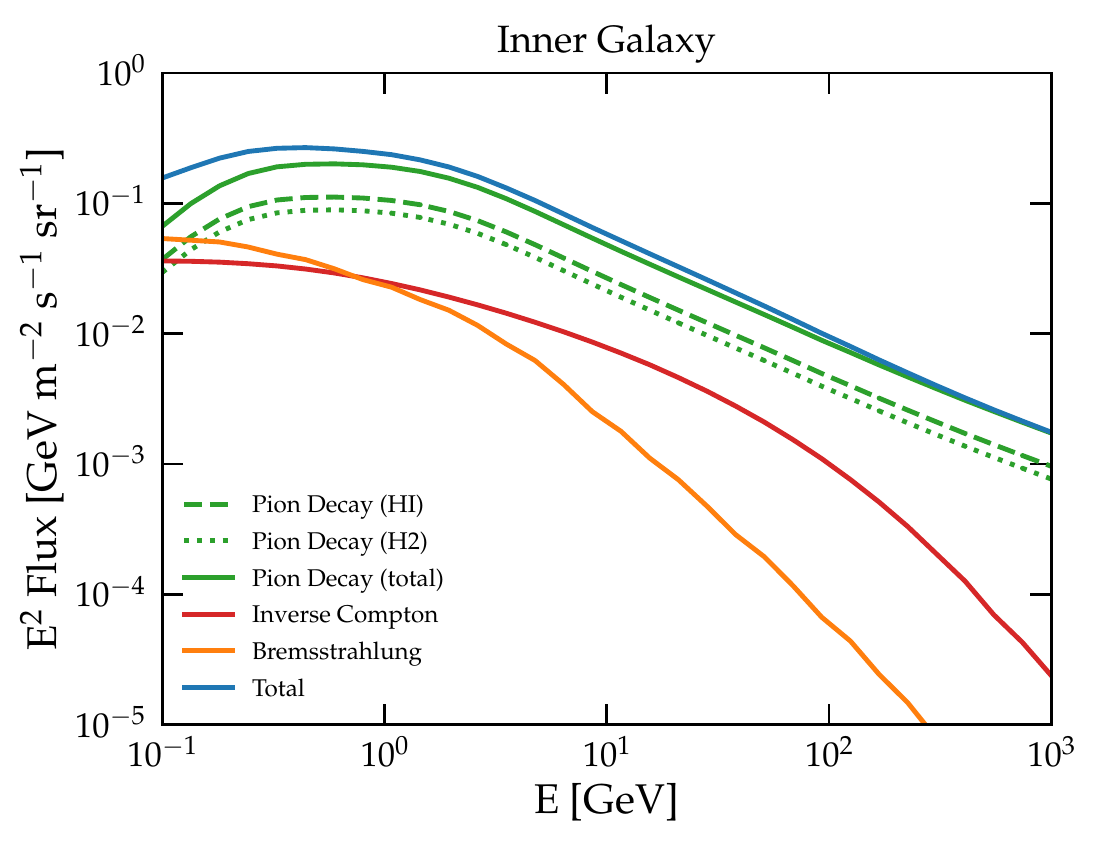}
\includegraphics[width=0.9\columnwidth]{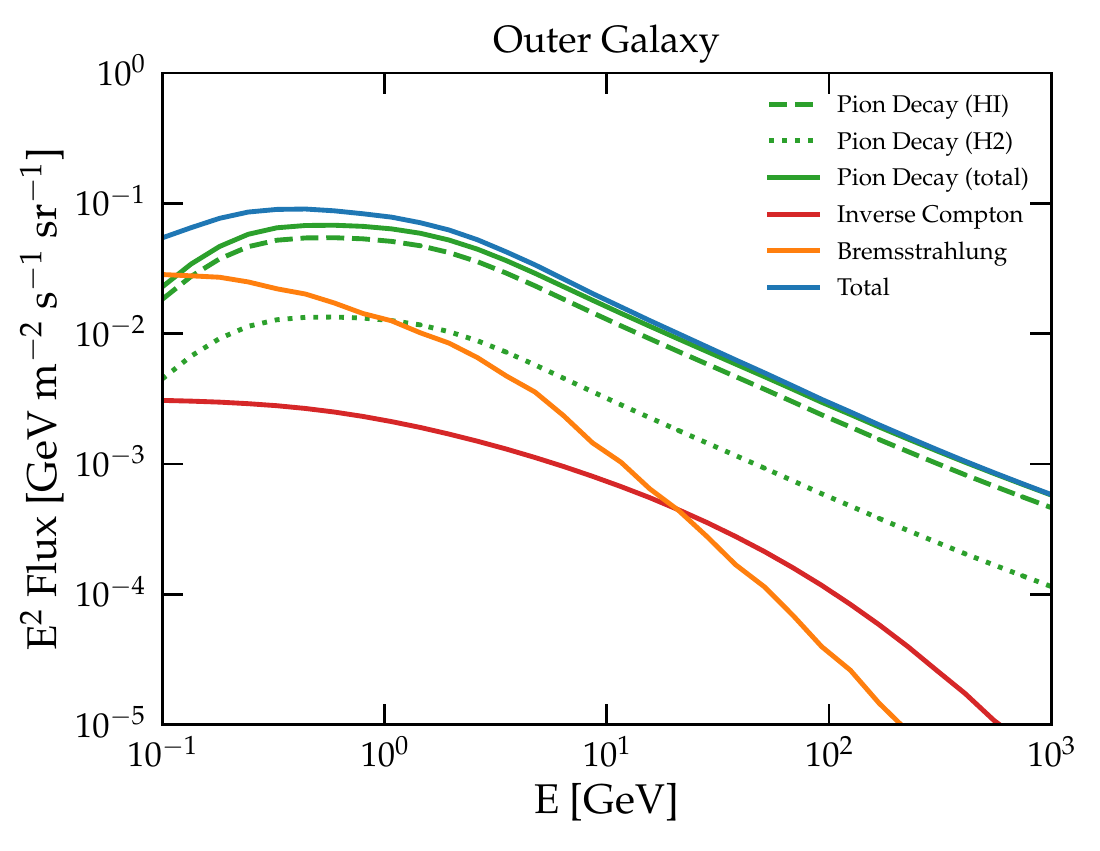}
\caption{Gamma-ray spectra extracted from four sky regions: full-sky (top-left), intermediate latitudes ($10^\circ < | b | < 40^\circ$, top-right), inner Galaxy ($|b|<8^\circ \wedge 80^\circ<l<280^\circ$, bottom-left), and outer Galaxy ($|b|<8^\circ \wedge l < 80^\circ \wedge l > 280^\circ$, bottom-right). The total gamma-ray spectrum (blue solid line) is computed as the sum of three basic emission components: inverse Compton (red solid line), bremsstrahlung (orange solid line), and $\pi^0$-decay (green solid line). For the pion decay we show the contributions of HI (green dashed line) and of H$_2$ (green dotted line) separately.}
\label{fig:gammaray-spectra-GeV}
\end{figure*}

In this section we analyze the \HERMES~predictions for the gamma-ray sky in the GeV-TeV domain.
As mentioned already in Section~\ref{sec:physics}, the main emission mechanisms in this energy range are $\pi^0$ decay, inverse Compton scattering, and bremsstrahlung.

Computation of the emission from $\pi^0$ decay is required when modeling the target gas distribution as a set of column density full-sky maps associated to different Galactocentric rings as described in Section~\ref{sec:piondecay}.
In \HERMES,~we adopt the recent model developed by Q.~Remy of the molecular and atomic interstellar gas~\citep{Galview1}.

In particular the molecular gas distribution is based on the $115$ GHz CfA CO survey in
~\cite{Dame2001apj} and \cite{Dame2004aspc}, which provides a high-resolution ($0.125^\circ$ spacing) sampling  along the Galactic plane, and a slightly less refined sampling ($0.25^\circ$) at high latitudes ($|b| > 10^\circ$). 
Concerning the atomic part, the model adopts the HI4PI survey by~\cite{HI4PI2016} of the 21-cm emission of neutral hydrogen (the angular resolution is $\simeq 0.26^\circ$), which traces the whole neutral atomic gas, including the cold (CNM) and the warm (WNM) components.
The emission from each line of sight is decomposed into $11$ different Galactocentric rings by taking into account the Doppler shift from the Galactic rotation. The rotation curve adopted for this purpose is the one derived in~\cite{Sofue2015pasj}. For the molecular component, two dedicated rings for the inner and outer regions of the central molecular zone are also present.

For this process, the gamma-ray production cross-section is based on the model developed in \cite{Kamae2006apj}. 
The Kamae formulation entails a parametrization of the gamma-ray, neutrino, and secondary leptons produced by proton-proton interactions in the ISM featuring a logarithmically rising inelastic cross-section, a description of the diffraction dissociation process, and the Feynman scaling violation. The model is validated for CR proton energy up to $\simeq 500$ TeV.

For the IC calculation, we adopt the ISRF model from~\cite{Vernetto2016prd}, which includes three main components: the uniform cosmic microwave background (CMB), and the spatially dependent infrared (radiated by interstellar dust heated by stars) and stellar emissions (see also~\citealt{Evoli2017jcap}). 

Finally, the bremsstrahlung component is computed with the same gas distribution as for the calculation of the $\pi^0$ emissivity, and with the differential cross-section described in~\cite{Tsai1974rvmp}.
Figure~\ref{fig:gammaray-spectra-GeV}  shows the resulting spectra in the GeV - TeV range associated to the different emission components in the following  regions of interest (ROIs): full sky; inner Galaxy: $-80^{\circ} < l < 80^{\circ}$; $8^{\circ} < b < 8^{\circ}$; outer Galaxy: $|l| > 80^{\circ}$; $8 < b < 8^{\circ}$; intermediate latitudes: $10^{\circ} < b < 40^{\circ}$.


We notice that the emission from neutral pion decay dominates above $\sim 1$~GeV in all the ROIs we considered.
The main contribution in the inner Galaxy is due to the interactions with the molecular hydrogen, which is mostly concentrated in a relatively narrow strip (scale height $\lesssim 70$ pc) along the Galactic plane, while at intermediate latitudes the atomic component plays the dominant role due to its more extended latitude profile.

The leptonic IC emission is slightly harder in the $1$ -$ 10$ GeV range, and exhibits a clear cutoff at the highest energies. 
This feature stems from the cutoff in the parent electron population, which is assumed here at $1$ TeV, while the proton population responsible for the $\pi^0$ emission is modeled as an almost featureless power law up to the CR knee (with the only exception being the mild hardening at a rigidity $\simeq 200$ GV as identified by the PAMELA and AMS experiments).
The IC emission always appears subdominant with respect to the hadronic component, but it becomes more relevant in the mid-latitude regions, because of the larger scale height of the ISRF (compared to both the molecular and atomic gas).

Figures~\ref{fig:gammaray-skymap-pi0-HI-10GeV} and~\ref{fig:gammaray-skymap-IC-10GeV} show the full-sky maps associated to the aforementioned components. The $\pi^0$ intensity maps are shown separately for the neutral and the molecular hydrogen contributions. 
The high level of detail of the hadronic maps reflects the high resolution of the radio observations tracing both molecular and atomic gas. 
Several patterns can be clearly recognized in the maps. The Galactic plane shines in gamma rays, exhibiting a flux that is three orders of magnitude higher than the polar regions; the North Polar Spur, the brightest part of Loop I, is also clearly visible above the plane, similarly to in the radio map. 
At $l \simeq 70^{\circ}$ along the plane, some small-scale features associated to the Cygnus region are also evident, in particular in the map related to the molecular target.   
We also notice the different level of clumpiness and spatial extension between the maps of the molecular and the neutral hydrogen, with the latter being more spread out and diffuse along the Milky Way.
On the other hand, the IC map appears remarkably smoother, because it reflects the more homogeneous distribution of the target.

\begin{figure*}
\centering
\includegraphics[width=0.9\columnwidth]{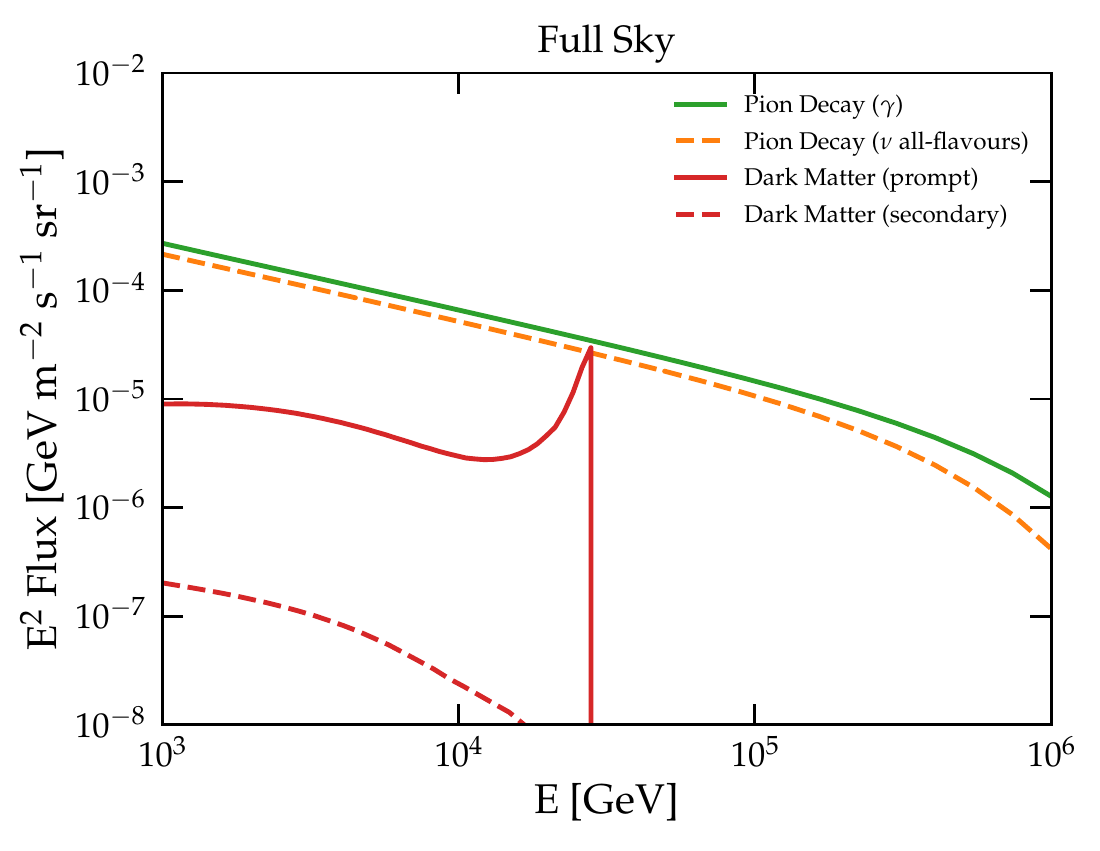}
\includegraphics[width=0.9\columnwidth]{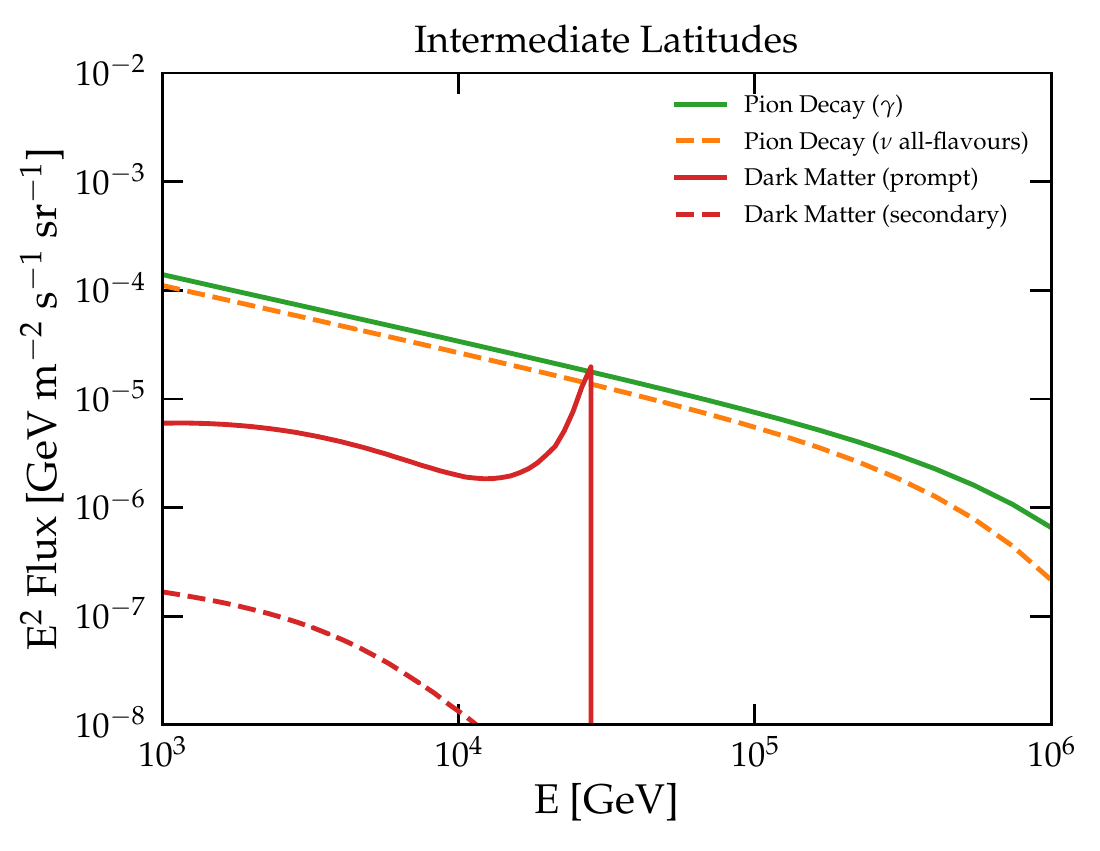}
\includegraphics[width=0.9\columnwidth]{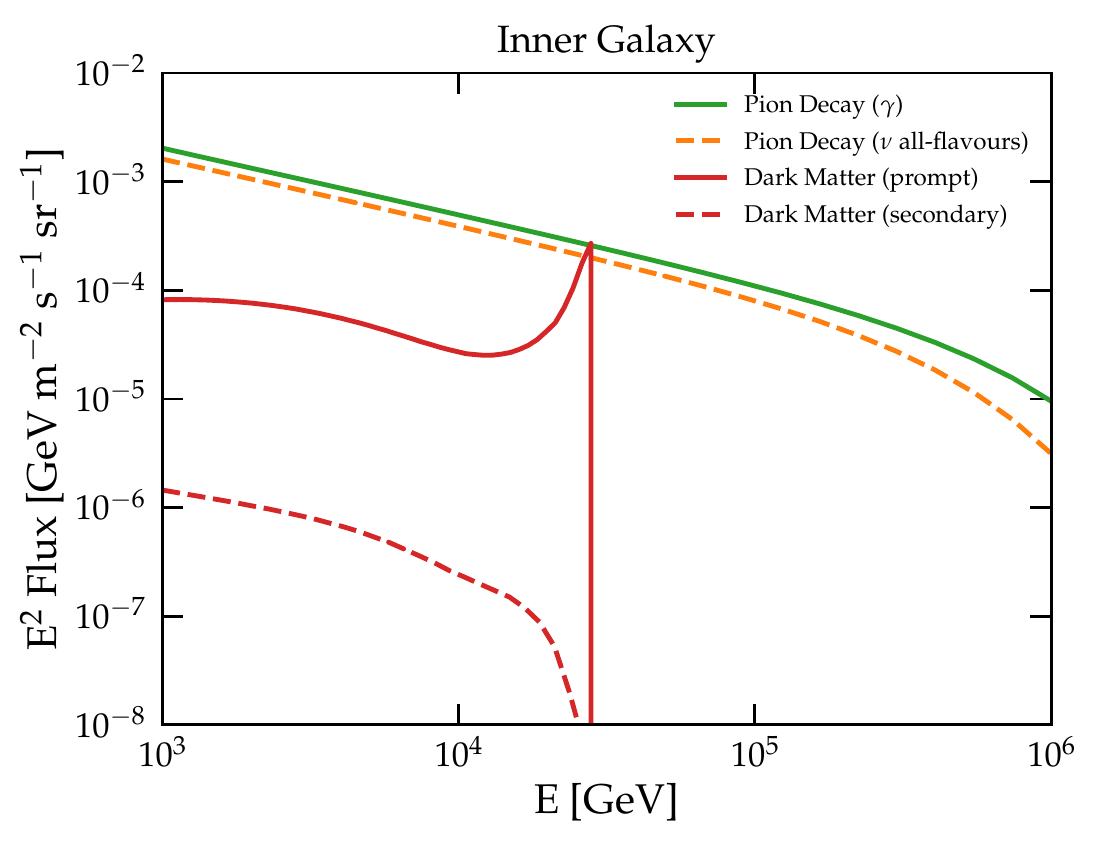}
\includegraphics[width=0.9\columnwidth]{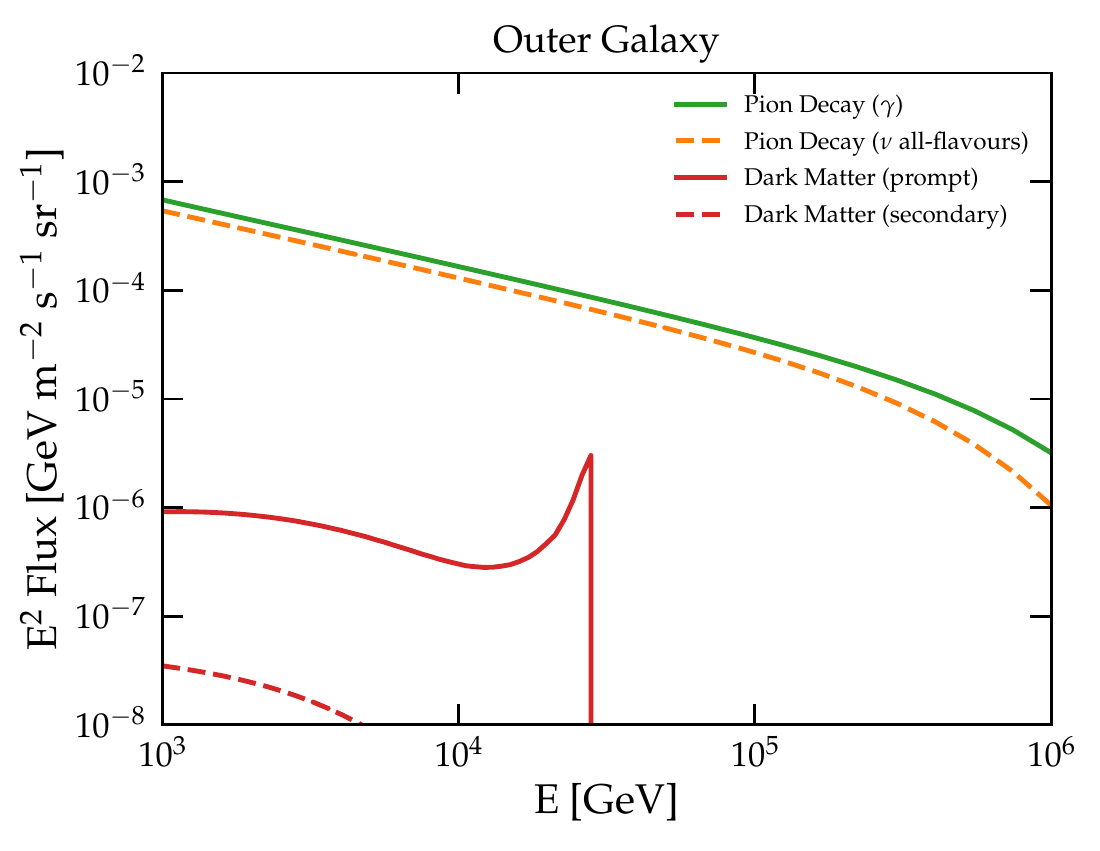}
\caption{Spectra extracted from the same four sky regions as in Fig.~\ref{fig:gammaray-spectra-GeV}. The gamma-ray spectrum from the $\pi^0$-decay (green solid line) is compared with the $\nu$ spectrum (summed over all flavors) attributable to $\pi^\pm$-decay (orange dashed line). 
The prompt gamma-ray spectrum expected from DM annihilation in $W^+W^-$ (with mass $M_\chi = 30$~TeV and cross-section $\langle \sigma v \rangle = 3\times 10^{27}$~cm$^2$/s) is shown with a red solid line. The IC from DM secondary leptons is also shown with a red dashed line.}.
\label{fig:gammaray-spectra-TeV}
\end{figure*}

\begin{figure}
\centering
\includegraphics[width=0.93\columnwidth]{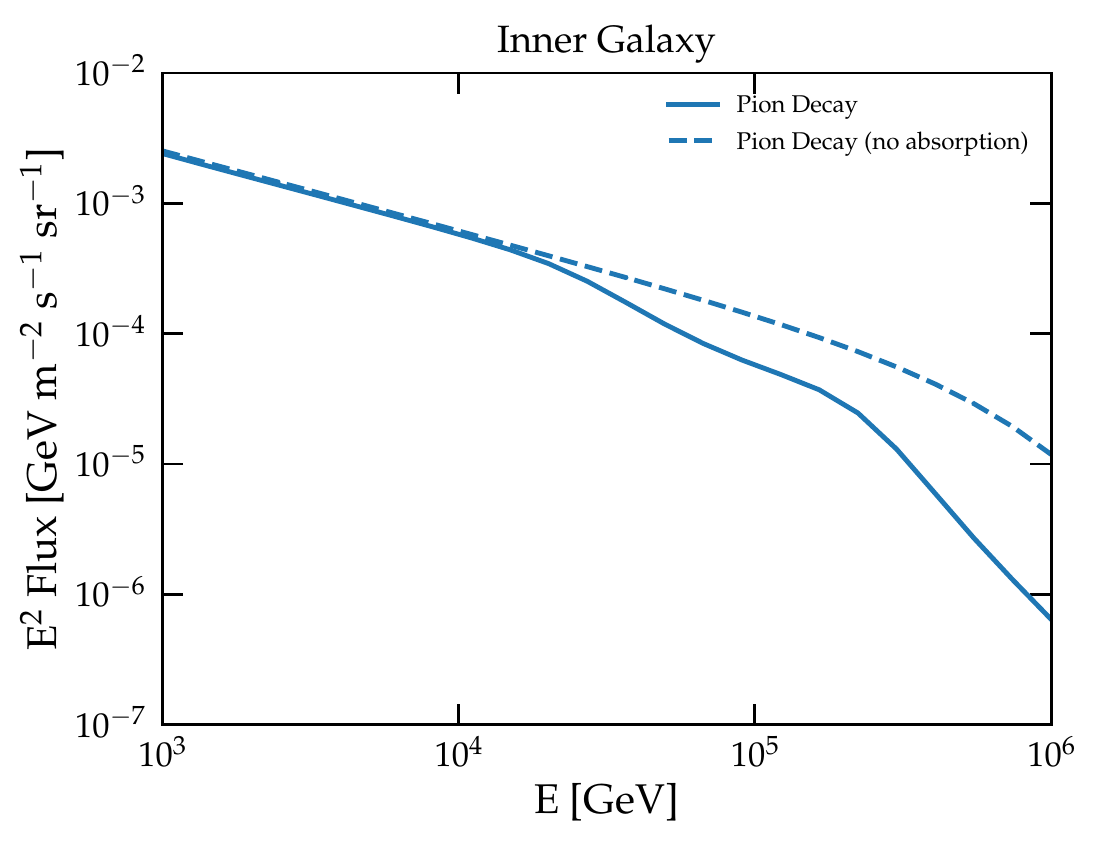}
\includegraphics[width=0.93\columnwidth]{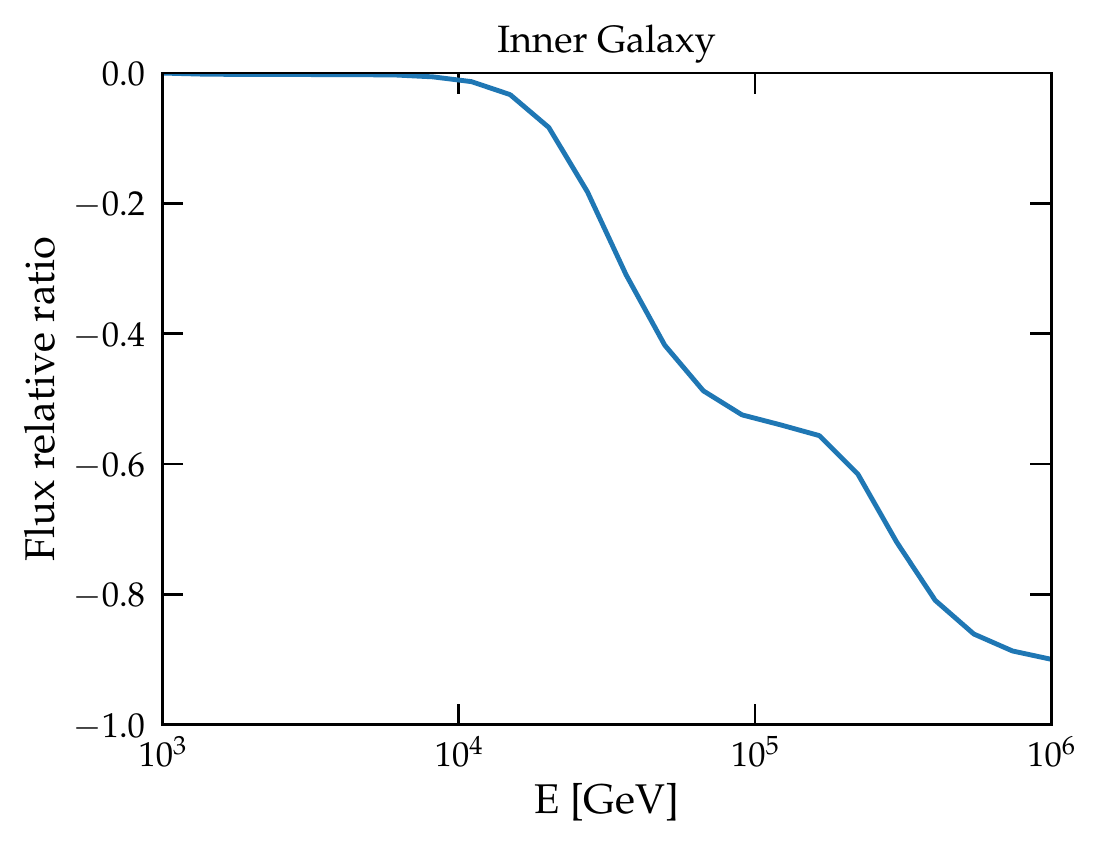}
\caption{Top panel: Spectrum of the $\pi^0$ emission computed for the inner Galaxy ROI (as defined in Fig. \ref{fig:gammaray-spectra-GeV}) together with the case without absorption. Bottom panel: Relative ratio of the flux between the case without and with absorption.}
\label{fig:gammaray-spectra-absorption}
\end{figure}

The results shown in this section provide a conservative estimate ({\it base model}) of the nonthermal emission from the Galaxy, namely from radio frequencies all the way up to multi-TeV gamma rays, based on our current knowledge of the normalization and the spectrum of local CR fluxes, of the distribution of the astrophysical targets (as discussed above), and of the simplest assumptions on CR transport. 

A comparison or refitting to actual radio and gamma-ray data is beyond the scope of this paper, and so the model should be considered as the minimal conservative {prediction} of the nonthermal emission from the Milky Way based on the simplest theoretical arguments and independent from any nonlocal observable.

\subsection{High-energy Gamma-ray sky}
    
Here we show the capability of \HERMES~in the context of the multi-TeV energy domain.
We adopt the same models for the astrophysical targets as described in the previous section. As far as cross sections are concerned, for the gamma-ray and neutrino production we implement the model described in~\cite{Kelner2008prd} (with the parameterization of the proton--proton total inelastic cross-section updated as in~\citealt{Kafexhiu2014prd}).

The features of the code we want to emphasize in this context are: (i) the calculation of gamma-ray absorption which becomes relevant above $\sim 10$~TeV; (ii) the consistent modeling of different messengers (gamma-rays and neutrinos); and (iii) the calculation of DM-induced gamma-ray and neutrino signals, which is of particular relevance because indirect searches for multi-TeV DM candidates will be highly abundant in the coming years~\citep{CTA-DM2021}.

Figure~\ref{fig:gammaray-spectra-TeV} shows the spectra of gamma-ray (without absorption) and neutrino fluxes as predicted for the pion-decay process, and we compare them with the signal expected for a specific DM candidate.
As expected, the neutrino (summed over all flavors) and the gamma-ray spectra exhibit a similar spectral slope and comparable normalization. At the highest energies, the two spectra behave slightly differently because of kinematic effects. 
The $\nu/\gamma$ ratio is smaller than 1 at these energies as also found in previous calculations (see for instance the tables reported in \citealt{Cavasinni2006aph}).

To illustrate the DM case, we consider a WIMP with $M_\chi = 30$ TeV mainly annihilating into $W^+ W^-$ with a cross-section ten times larger than the thermal reference cross-section.
Such a combination of mass, annihilation channel, and cross-section is within the reach of the expected performances of CTA.
The DM particles are distributed in the Galaxy with a gNFW profile (see Section~\ref{sec:dm}). The fluxes at production (normalized to one annihilation) for secondary leptons and prompt gamma rays are taken from the public repository\footnote{available at \url{http://www.marcocirelli.net/PPPC4DMID.html}}~(see~\citealt{Cirelli2011jcap,Ciafaloni2011jcap}). 

In Fig.~\ref{fig:gammaray-spectra-TeV} we compare the signal expected by the prompt emission with the one by secondary emission. 
The latter is associated to the diffuse population of electron+positron pairs produced as a consequence of the W-boson decays. 
We model the propagation of the secondary particles with \DRAGON~and compute the corresponding IC emission with~\HERMES.
We notice that, in this energy range, the signal is dominated by the prompt emission. This component features a broad soft continuum due to the fast decay of neutral pions produced by the hadronization of the quark+antiquark pairs in the final state, and a hard feature at high energy with a sharp cutoff associated to final-state radiation. 
This spectral feature makes the DM signal comparable in intensity with the smooth astrophysical flux at energies close to the DM mass, enabling future searches in the inner Galaxy region with high-resolution observations by upcoming Cerenkov telescopes. 

Finally, the impact of gamma-ray absorption is shown for the inner Galaxy in Fig.~\ref{fig:gammaray-spectra-absorption}.
The difference becomes relevant starting at $\sim 10$ TeV and a suppression as large as $\simeq 60$\% is reached at $\simeq 100$ TeV. 
Including the absorption is therefore relevant for predictions of the very-high-energy flux measured by dedicated experiments such as HAWC and LHAASO.

\section{Conclusions}
\label{sec:conclusions}

We present the \HERMES~public numerical package which is designed to generate simulated maps and spectra of radio, gamma-ray, and neutrino diffuse emissions originating from the interactions of the galactic CR population with the interstellar environment or from DM annihilations and decays in the halo.
The physical processes implemented in the code are discussed in Section~\ref{sec:physics}.

The modular structure, described in Section~\ref{sec:code}, enables the user to combine independent modules to study multiple use cases and at the same time enables simple updates of all models to meet the current needs of the users.

In order to demonstrate the full capabilities of \HERMES~we provide example sky maps and spectra (Section~\ref{sec:applications}) computed using up-to-date input models from the literature, including the simulated spectra of the prompt and secondary diffuse gamma-ray emission due to the annihilation of a realistic DM particle physics candidate. 
A run configuration, including the ingredients to compute the map and the resolution parameters, is fully specified by a Python or C++ script file to be compiled linking \HERMES~as an external library. To show this, we provide a full and detailed example of a Python script that can be used to compute the gamma-ray sky map of the $\pi^0$-decay process.

\HERMES~can be used to constrain the properties of the Galactic CR population and increase our understanding of the radio and gamma-ray diffuse Galactic emission. 

\section*{Acknowledgments}

We are grateful to Pierre Cristofari for reading the text thoroughly and giving helpful feedback. We thank Tess Jaffe, Ralf Kissmann, Quentin Remy, Andy Strong, Luigi Tibaldo and Silvia Vernetto for useful conversations and insights. We further thank the {\tt CRPropa} development team for providing a good role model for developing a high-quality C++ astrophysical code and making it available as free software, aside from the fact that \HERMES~borrows  {\tt CRPropa}'s implementation of magnetic fields and abstract vector and grid classes. A.D.~acknowledges the P.O.I.N.T. Association, Kri\v{z}evci, Croatia for providing computational resources needed for testing~\HERMES.

This work was funded through Grants ASI/INAF No. 2017-14-H.O.

D.~Gaggero has received financial support through the Postdoctoral Junior Leader Fellowship Programme from la Caixa Banking Foundation (grant n.~LCF/BQ/LI18/11630014). D.~Gaggero was also supported by the Spanish Agencia Estatal de Investigaci\'{o}n through the grants PGC2018-095161-B-I00, IFT Centro de Excelencia Severo Ochoa SEV-2016-0597, and Red Consolider MultiDark FPA2017-90566-REDC. 

C.~Evoli acknowledges the European Commission for support under the H2020-MSCA-IF-2016 action, Grant No. 751311 GRAPES 8211 Galactic cosmic RAy Propagation: An Extensive Study.

\bibliographystyle{aa}     
\bibliography{2021-hermes,software} 

\end{document}